\documentclass[]{emulateapj}
\PassOptionsToPackage{hyphens}{url}\usepackage{hyperref}
\usepackage{natbib}
\usepackage{amssymb}
\usepackage{color}
\usepackage{amsmath,mathtools}
\usepackage{epsfig}
\usepackage[FIGTOPCAP]{subfigure}
\usepackage{afterpage}
\usepackage{enumerate}
\usepackage{multirow}
\usepackage{verbatim}
\usepackage{relsize}
\usepackage{tikz}
\usetikzlibrary{shapes.geometric, arrows}
\usepackage{morefloats}
\usepackage{wasysym}

\newcommand{\noop}[1]{}

\shorttitle{EVEREST}
\shortauthors{Luger et al. 2016}

\begin{document}

\title{EVEREST: Pixel Level Decorrelation of \emph{K2} Light curves}

\author{Rodrigo Luger\altaffilmark{1,2}, Eric Agol\altaffilmark{1,2}, Ethan Kruse\altaffilmark{1}, Rory Barnes\altaffilmark{1,2},\\
Andrew Becker\altaffilmark{1}, Daniel Foreman-Mackey\altaffilmark{1,4}, and Drake Deming\altaffilmark{3}}
\altaffiltext{1}{Astronomy Department, University of Washington, Box 351580, Seattle, WA 98195, USA; \href{mailto:rodluger@uw.edu}{rodluger@uw.edu}}
\altaffiltext{2}{Virtual Planetary Laboratory, Seattle, WA 98195, USA}
\altaffiltext{3}{Department of Astronomy, University of Maryland, College Park, MD 20742, USA}
\altaffiltext{4}{Sagan Fellow}
\begin{abstract}
We present \texttt{EVEREST}, an open-source pipeline for removing instrumental noise
from \emph{K2} light curves. \texttt{EVEREST} employs a variant of pixel level decorrelation (PLD) 
to remove systematics introduced by the spacecraft's 
pointing error and a Gaussian process (GP) to capture astrophysical variability. We
apply \texttt{EVEREST} to all \emph{K2} targets in campaigns 0-7, yielding light curves
with precision comparable to that of the original \emph{Kepler} mission for stars brighter
than $K_p \approx 13$, and within a factor of two of the \emph{Kepler} precision for fainter
targets. We perform cross-validation and transit injection and recovery tests to validate
the pipeline, and compare our light curves to the other de-trended light curves available
for download at the MAST High Level Science Products archive. We find that 
\texttt{EVEREST} achieves the highest average precision of any of these pipelines for 
unsaturated \emph{K2} stars. The improved precision of these
light curves will aid in exoplanet detection and characterization, investigations of stellar
variability, asteroseismology, and other photometric studies. The \texttt{EVEREST} pipeline can
also easily be applied to future surveys, such as the \emph{TESS} mission, to
correct for instrumental systematics and enable the detection of low signal-to-noise transiting
exoplanets. The \texttt{EVEREST} light curves and the source code
used to generate them are freely available online.
\end{abstract}

\section{Introduction}
\label{sec:intro}
Launched in 2009, the \emph{Kepler} spacecraft has to date led to the discovery
of nearly 5000 extrasolar planet candidates
and to a revolution in several fields of astronomy including but not limited to
exoplanet science, eclipsing binary characterization, asteroseismology and stellar variability studies.
Its unprecedented photometric precision allowed for the study of astrophysical signals
down to the level of $\sim 15$ parts per million \citep{GIL11}, which has enabled the discovery
of small planets in the habitable zones of their stars \citep[e.g.,][]{BOR13,QUI14,TOR15}.
Unfortunately, after the failure of its second reaction wheel in May 2013, the spacecraft was
no longer able to achieve the fine pointing accuracy required for high precision photometry,
and the nominal mission was brought to an end. Engineering tests suggested that by aligning
the spacecraft along the plane of the ecliptic, pointing drift could be mitigated
by the solar wind pressure and by periodic thruster firings. As of May 2014 the spacecraft has been operating in
this new mode, known as \emph{K2}, and has continued to enable high precision photometry
science, monitoring tens of thousands of stars near the plane of the ecliptic during campaigns 
lasting about 75 days each \citep{HOW14}.

However, because of the reduced pointing accuracy, \emph{K2} raw aperture photometry is between 
3 and 4 times less precise than that of the original \emph{Kepler} mission and displays
strong instrumental artefacts with different timescales, including a $\sim$ 6 hour trend, which
severely compromise its ability to detect small transits. Recently, several authors
have developed powerful methods to correct for these systematics, often coming within
a factor of $\sim$ 2 of the original \emph{Kepler} precision.

In particular, the \texttt{K2SFF} pipeline \citep{VJ14} decorrelates \emph{K2} aperture 
photometry with the centroid position
of the stellar images. Centroids are determined based either on the center-of-light
or via a Gaussian fit to the stellar PSF. The motion of the centroids is then fit with a polynomial
and transformed into a single parameter that relates spacecraft motion to flux variations,
which is then used to de-trend the data. Similar methods are employed in the \texttt{K2VARCAT} pipeline \citep{ARM15}, 
developed specifically for variable \emph{K2} stars, the \texttt{K2P$^2$} pipeline \citep{LUN15},
which uses an intelligent clustering algorithm to define custom apertures, and in the pipeline
of \cite{HUA15}, which employs astrometric solutions to the motion of \emph{K2} targets,
determining the $X$ and $Y$ motion of a target from the behavior of multiple stars on
the same spacecraft module. Finally, the \texttt{K2SC} pipeline \citep{AIG15,AIG16} and 
the pipeline of \cite{CRO15} both employ a Gaussian process (GP) to remove
instrumental noise, using the $X$ and $Y$ coordinates of the target star as the regressors to derive
a model for the instrumental systematics. The nonparametric nature of the GP results 
in a flexible model with increased de-trending power especially for dim \emph{K2}
targets.

In one way or another, all of these techniques rely on numerical methods to identify and 
remove correlations between the stellar position and the intensity fluctuations. Even
when a nonparametric technique such as a GP is used, assumptions are still made about
the nature of the correlations between spacecraft motion and instrumental variability.
Moreover, the process of determining the stellar centroids is prone to uncertainties
and relies on assumptions about the shape of the stellar PSF.

A powerful alternative to these methods is pixel level decorrelation (PLD), a method developed
by \cite{DEM15} to correct for systematics in \emph{Spitzer} observations of transiting
hot Jupiters. The tenet of PLD is that the best way to correct for noise introduced by
the motion of the stellar image does not involve actual measurements of the position
of the star. The centroid of the stellar image is, after all, a secondary data product
of photometry, and is subject to additional uncertainty.
PLD skips these two numerical steps (i.e., fitting for the stellar position and solving
for the correlations) by operating on the \emph{primary} data products of photometry, the
intensities in each of the detector pixels. 
These intensities are normalized by the
total flux in the chosen aperture then used as basis vectors for a linear least-squares 
(LLS) fit to the aperture-summed flux. Since astrophysical signals (stellar variability,
planet transits, stellar eclipses, etc.) are present in all of the pixels in the aperture,
the normalization step removes astrophysical information from the basis set, ensuring
that PLD is sensitive only to the signals that are \emph{different} across the aperture. PLD is
therefore an ``agnostic'' method of performing robust flat-fielding corrections with
minimal assumptions about either the nature of the intra-pixel variability or the correlation 
between spacecraft jitter and intensity fluctuations. We note that our method is similar 
to that of \cite{DFM15} and \cite{MON15}, who use the principal components of the variability 
among the full set of \emph{K2} campaign 1 light curves as ``eigen light curve'' regressors.
However, rather than deriving our basis vectors from other stars in the field, whose
light curves contain undesired astrophysical signals, our basis vectors are derived
solely from the pixels of the star under consideration.

In this paper we build on the PLD method of \cite{DEM15}, extending it to higher order
in the pixel fluxes and performing principal component analysis (PCA) on the basis
vectors to limit the flexibility of the model and thus prevent overfitting. We further 
couple PLD with a GP to 
disentangle astrophysical and instrumental variability. 

We apply our pipeline, \texttt{EVEREST}
(\emph{EPIC Variability Extraction and Removal for
Exoplanet Science Targets}), to the entire set of \emph{K2} light curves from campaigns 0-7 and generate a 
publicly-available database of processed light curves. Our code is open-source and will be made
available online.

The paper is organized as follows: in \S\ref{sec:pld} we review the basics of PLD and
derive our third-order PLD/PCA/GP model, and in \S\ref{sec:methods} we describe our pipeline
in detail. Results are presented in \S\ref{sec:results}, and in \S\ref{sec:conclusions}
we conclude and outline plans for future work.

\section{Pixel Level Decorrelation}
\label{sec:pld}
\subsection{First Order PLD}
\label{sec:firstorder}
The linear PLD model developed in \cite{DEM15} is given by the expression
\begin{align}
\label{eq:demingmodel}
m_i = &\sum\limits_{l}a_l\frac{p_{il}}{\sum\limits_{k}p_{ik}} + \alpha + \beta t_i + \gamma t_i^2
\end{align}
where $m_i$ denotes the noise model at time $t_i$,
$p_{il}$ denotes the flux in the $l^{th}$ pixel at time $t_i$, and $a_l$ is
the linear PLD coefficient for the $l^{th}$ pixel. Both sums are taken over
all pixels in the aperture; the last three terms are a polynomial in time used
to capture temporal variations in the flux baseline due to intrinsic variability
of the star.

The coefficients are obtained by minimizing the sum of the squares of the difference
between the model $m_i$ and the simple aperture photometry (SAP) flux, $y_i$:
\begin{align}
\label{eq:demingsolution}
\frac{\partial \chi^2}{\partial a_l} = 0
\end{align}
where
\begin{align}
\label{eq:demingchisq}
\chi^2 = \mathlarger{\mathlarger{\sum\limits_{i}}}\frac{\left(y_i - m_i\right)^2}{\sigma_i^2}.
\end{align}
In the expression above, $\sigma_i$ are the standard errors of the flux and
\begin{align}
\label{eq:demingsap}
y_i = \sum\limits_{k}p_{ik}.
\end{align}

Framed in this manner, computation of the PLD model is a linear regression problem; the
coefficients are readily found by simultaneously solving Equation~(\ref{eq:demingsolution})
for all coefficients $a_l$, as well as $\alpha$, $\beta$, and $\gamma$.
%

\subsection{Higher Order PLD}
\label{sec:higherorder}
\cite{DEM15} found that the pointing jitter of the \emph{Spitzer} telescope was 
sufficiently small that extending PLD to higher order in the pixel fluxes was
unnecessary. However, because of the large pointing variations of the \emph{K2}
spacecraft on short timescales, we find it necessary to extend PLD to higher order.
Keeping terms up to third order in the pixel fluxes, we may express this model as 
\begin{align}
\label{eq:pldmodel}
m_i = &\sum\limits_{l}a_l\frac{p_{il}}{\sum\limits_{k}p_{ik}} + \nonumber\\
      &\sum\limits_{l}\sum\limits_{m}b_{lm}\frac{p_{il}p_{im}}{(\sum\limits_{k}p_{ik})^2} + \nonumber\\
      &\sum\limits_{l}\sum\limits_{m}\sum\limits_{n}c_{lmn}\frac{p_{il}p_{im}p_{in}}{(\sum\limits_{k}p_{ik})^3} + \alpha + \beta t_i + \gamma t_i^2
\end{align}
where once again index $i$ denotes time and all other indices denote the pixel number. 
The PLD coefficients are now $a_l$ (one per pixel), 
$b_{lm}$ (one per pixel pair), and $c_{lmn}$ (one per group of three pixels). Despite
the added complexity, the model remains linear and may be solved in a similar fashion 
as before.

\begin{figure}[h]
  \begin{center}
      \psfig{file=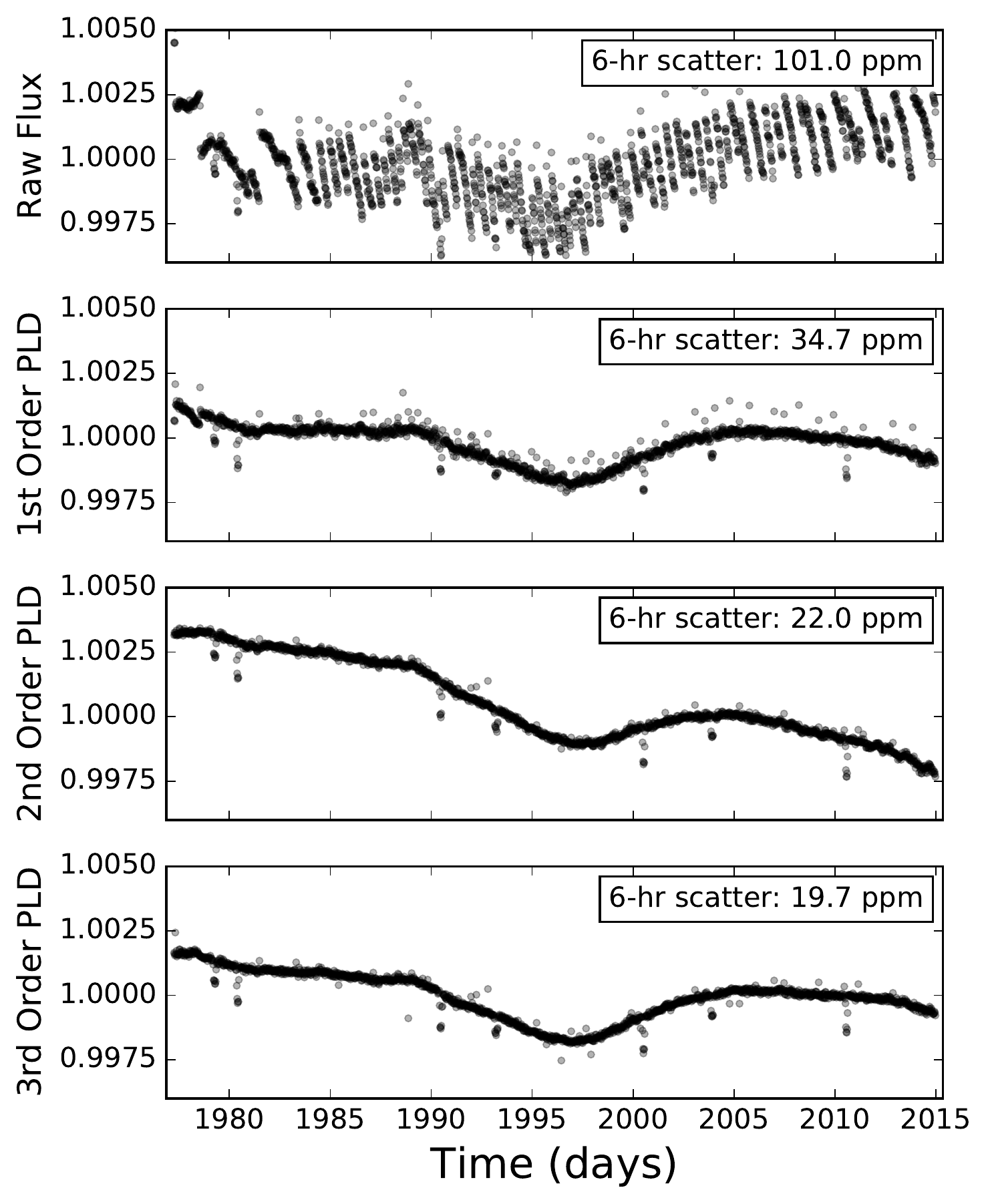,width=3.5in}
       \caption{PLD applied to a portion of the data for EPIC 201367065 (K2-3). 
       The top panel is the background-subtracted, normalized SAP flux in a large 
       35-pixel aperture centered on the target. The bottom three panels show 
       the normalized PLD-de-trended flux for 1$^{\mathrm{st}}$, 2$^{\mathrm{nd}}$, and 
       3$^{\mathrm{rd}}$ order PLD, respectively, using only the 10 brightest pixels. 
       PLD increases the 6-hr photometric precision by factors of 2.9 (1$^{\mathrm{st}}$ order), 4.7 
       (2$^{\mathrm{nd}}$ order), and 5.2 (3$^{\mathrm{rd}}$ order).}
     \label{fig:third_order}
  \end{center}
\end{figure}

In Figure~\ref{fig:third_order}, we illustrate the de-trending technique for EPIC
201367065 (K2-3), a star host to three known transiting planets \citep{CRO15}. The top
panel shows the normalized raw aperture-summed flux after subtracting off the 
background; note the large systematics at very short ($\sim$ 6 hr) timescales. The
next panel shows the flux de-trended with first order PLD (Equation~\ref{eq:demingmodel}).
The scatter is reduced by a factor of 2.9 and the
seven transits become visible by eye. The following two panels show the results of
second and third order PLD, which improve the scatter over the raw data by factors of 
4.7 and 5.2, respectively. Note, importantly, that even data collected during thruster
fire events (the outlier points seen above the continuum every $\sim$ 6 hr in the top 
two plots), is properly corrected by higher order PLD.

One might wonder why not go to even higher order in the pixel coefficients. While possible in principle,
the number of regressors increases steeply with the PLD order. For a typical \emph{K2} star, this number is
around $5\times 10^3$ for third order PLD, and would increase to $\sim5\times 10^4$ for
fourth order and $\sim3\times 10^5$ for fifth order PLD.
%
%
While such a large number of regressors can become computationally expensive, the 
most serious drawback of higher order PLD is that it can become so flexible as to
lead to overfitting. As we discuss in \S\ref{sec:dmopt} below,
PLD can sometimes overstep and remove part of the white noise component of the 
light curve. This is particularly problematic when the number of regressors becomes very large,
in which case PLD can lead to artificially low scatter in the de-trended light curve,
washing out astrophysical signals (including transits) in the process.

To avoid these issues, in Figure~\ref{fig:third_order} we computed the PLD basis vectors from
the 10 brightest pixels in the aperture. In practice, however, we obtain better results
by instead performing principal component analysis (PCA) on the first,
second, and third order fractional pixel fluxes (the terms in Equation~\ref{eq:pldmodel}).
This gives us a set of $N$ basis vectors $\mathbf{x}$ that describe most of the instrumental
variability in the light curve. We use these to construct our \emph{design matrix} 
\begin{align}
\label{eq:designmatrix}
\mathbf{X} =  \begin{pmatrix}
              x_{0,0} & x_{0,1} & ... & x_{0,N} & 1\\
              \\
              x_{1,0} & x_{1,1} & ... & x_{1,N} & 1\\
              \\
              ... & ... & ... & ... & ...\\
              \\
              x_{M,0} & x_{M,1} & ... & x_{M,N} & 1\\
              \end{pmatrix}
\end{align}
where $M$ denotes the total number of data points along the time dimension. We choose
the number of principal components $N$ that maximize the de-trending power while preventing
overfitting. We explain our method in \S\ref{sec:dmopt} below; typically, $N < 200$.

Since our problem is still linear, our model is simply
\begin{align}
\label{eq:pcamodel}
\mathbf{m} = \mathbf{X} \cdot \mathbf{c}
\end{align}
where $\mathbf{c}$ is the vector of coefficients, one for each basis vector in $\mathbf{X}$.
Their values are given by solving the generalized least-squares (GLS) problem
\begin{align}
\label{eq:pcacoefficients}
\mathbf{c} = \left(\mathbf{X}^\top \cdot \mathbf{K}^{-1} \cdot \mathbf{X}\right)^{-1} \cdot \left(\mathbf{X}^\top \cdot \mathbf{K}^{-1} \cdot \mathbf{y}\right)
\end{align}
where $\mathbf{K}$ is the covariance matrix of the data and $\mathbf{y}$ is the
SAP flux given by Equation~(\ref{eq:demingsap}). Note that for a diagonal covariance
$\mathbf{K}_{ij} = \delta_{ij}\sigma_i^2$, Equation~(\ref{eq:pcacoefficients}) is
mathematically equivalent to Equations~(\ref{eq:demingsolution}) and 
(\ref{eq:demingchisq}) from before.

\subsection{Gaussian Process Regression}
\label{sec:gp}
\begin{figure*}[t]
  \begin{center}
    \leavevmode
      \psfig{file=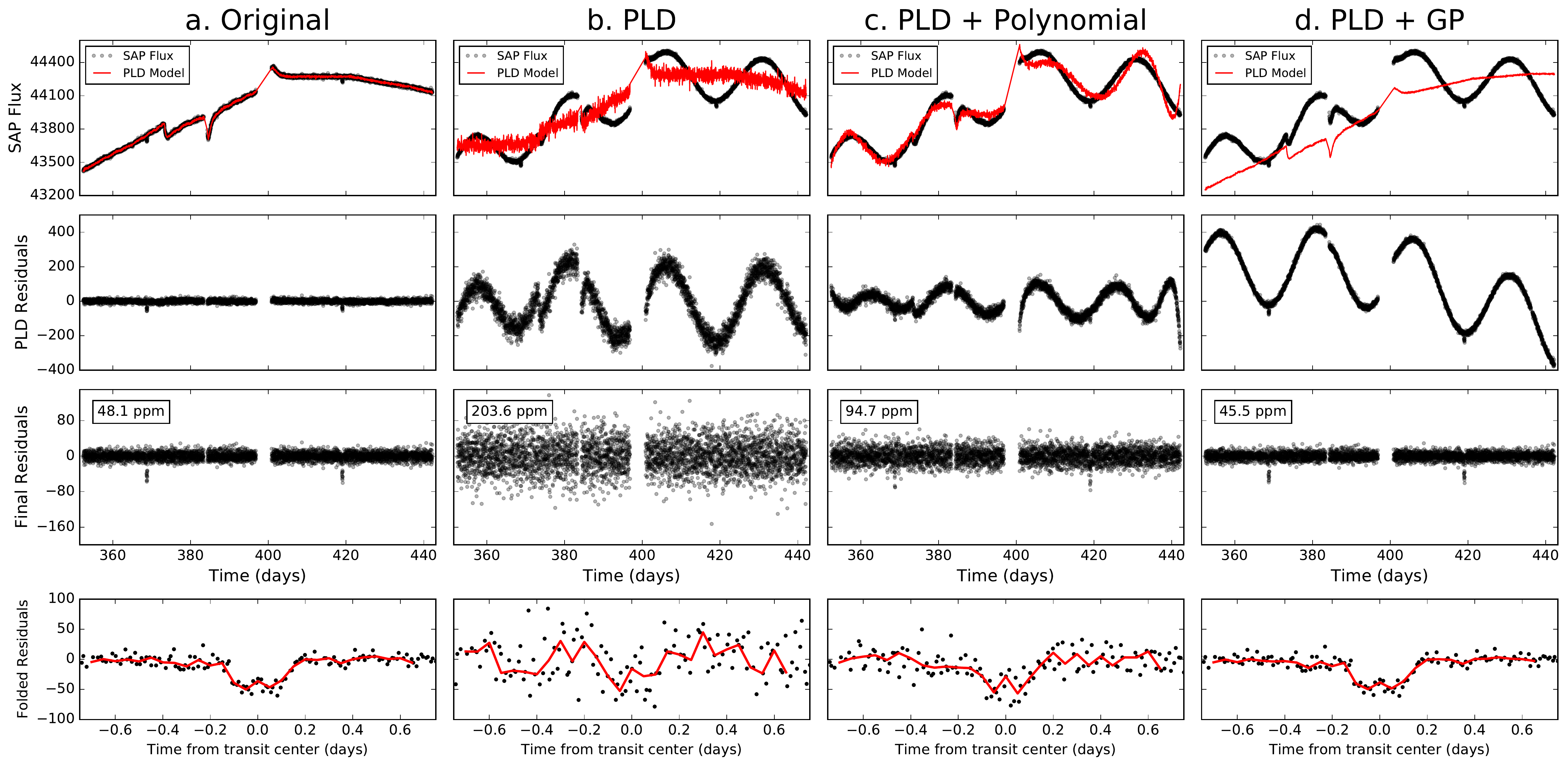,width=\textwidth}
       \caption{Different de-trending techniques for quarter 4 of KIC 8583696 (KOI 1275),
                a planet candidate host from the original \emph{Kepler} mission. The original
                data are shown in the left column; in the other columns we
                artificially injected a sinusoidal signal with a period
                of 25 days and an amplitude comparable to that of the instrumental variability.
                The top row shows the raw SAP data (black) and the first order
                PLD model (red); the residuals of the fit are indicated directly below. 
                The third row shows the final residuals after smoothing 
                with a GP to eliminate low-frequency stellar variability. Finally, the bottom
                row shows these residuals folded on the orbital period of the planet
                candidate (black), with the 1-hr median indicated in red. Combining PLD with a GP
                ensures PLD fits out only the instrumental variability without inflating the white noise.
                }
     \label{fig:sinusoid}
  \end{center}
\end{figure*}

In Equation~(\ref{eq:designmatrix}) we purposefully neglected the temporal polynomial
terms. In principle, modeling stellar variability should not be necessary if all we wish 
to remove is the instrumental noise. By virtue of using the
\emph{fractional} pixel fluxes as basis vectors (from which the astrophysical signal
has been removed by the normalization), PLD should fit out instrumental variability only,
obviating the need for an extra temporal term. However, in practice, this is not always
the case. To illustrate this, in Figure~\ref{fig:sinusoid} we plot a quarter of data
of KOI 1275, a star with one planet candidate discovered by the original \emph{Kepler}
mission \citep{BOR11}. We chose this star over one from the \emph{K2}
mission simply because it is easier to discriminate between instrumental and
astrophysical signals by eye for this light curve, though our arguments apply
equally well to \emph{K2}.

Figure~\ref{fig:sinusoid}\textbf{a} shows the raw simple aperture photometry (SAP) data in black (top
panel); in red we plot a simple first-order PLD fit with no polynomial term (obtained
from the solution to Equation~\ref{eq:demingsolution}). Below it are 
the residuals of the PLD fit (second panel, black) and the fully de-trended data after 
smoothing with a GP to remove stellar noise (third panel). The bottom panel shows 
the de-trended data folded on the period of the planet candidate, with the 1-hr 
median shown in red. Note that the PLD fit is quite good, as it removes the
low-frequency arc as well as the flux jump at $t \sim 375$ d and the thermal 
features at $t \sim 385$ d and $t \sim 400$ d. Since the star
isn't significantly variable, the temporal term is not necessary.

However, this is not the case when the star is variable. In the next three columns 
we multiply the pixel level light curve by a sinusoid 
with a period of 25 days to simulate stellar variability. We choose an amplitude 
comparable to the amplitude of the instrumental signal, and de-trend the light 
curve three different ways: PLD only (\textbf{b}), PLD plus a tenth order 
polynomial (\textbf{c}) and PLD plus a GP (\textbf{d}).

The PLD-only fit (Figure~\ref{fig:sinusoid}\textbf{b}) is quite poor. The general shape of the PLD
curve is mostly preserved and, as expected, the stellar signal is still present 
in the residuals, but the fit increases the white noise by a factor of $\sim$ 4,
all but washing out the transits (bottom panel). This happens because of a
serious limitation of the LLS problem framed in Equation~\ref{eq:demingsolution}, 
which yields the PLD coefficients that minimize the scatter of the PLD term $m_i$ 
about the flux $y_i$. If $m_i$ is not a suitable approximation to the flux, as in the case
of a highly variable star, the $\chi^2$ of the fit will necessarily be large.
It is not hard to show that $\chi^2$ can be substantially reduced
by inflating the white noise component of $m_i$, leading to the large scatter
seen in the de-trended data. Absent a model for the astrophysical variability,
PLD naturally exchanges correlated noise for white noise, which severely
compromises its de-trending power. 

A straightforward way to improve the quality of the fit is to include the polynomial
term to capture the stellar variability, as in \cite{DEM15}. We do this in Figure~\ref{fig:sinusoid}\textbf{c}, 
where we use a tenth order polynomial. While the fit is significantly
improved, PLD still inflates the white noise component, since the polynomial is 
not flexible enough to capture all of the stellar signal. Though the transit is
visible in the bottom plot, the quality of the de-trended data is significantly
degraded when compared to that of the non-variable star.

One might imagine that a higher order polynomial would do the trick. However, there
is an alternate approach, one that naturally follows from Equation~\ref{eq:pcacoefficients}.
Rather than model the stellar signal explicitly, we instead treat it non-parametrically
by performing Gaussian process regression, in
which we use a GP to estimate the covariance matrix $\mathbf{K}$. Provided
$\mathbf{K}$ is a reasonable approximation to the true covariance of the stellar signal,
Equation~\ref{eq:pcacoefficients} will yield a set of PLD coefficients that 
fit out only the instrumental component of the noise. We illustrate this in 
Figure~\ref{fig:sinusoid}\textbf{d}, where we use a Mat\'ern-3/2 kernel 
(see Equation~\ref{eq:kernels} below) with amplitude $\alpha_m = 100$ and timescale
$\tau_m = 20$ days to model the stellar signal. Unlike in the previous two cases,
the PLD model (top panel, red) no longer attempts to fit out the stellar signal; in 
fact, the curve is almost identical to the fit to the original data 
(Figure~\ref{fig:sinusoid}\textbf{a}). The astrophysical signal is
recovered to high fidelity by subtracting the PLD term (second panel). After removing it,
the final residuals (third panel) show a comparable (in fact, marginally smaller) RMS to the
residuals of the original data. It is also clear from the bottom panel that the transit 
shape and depth are well preserved.

We note that another advantage of GP regression is that the PLD model is
relatively insensitive to the particular choice of kernel function and values of its
hyperparameters. This can be seen in the example above; even though the stellar
variability signal was generated from a 25-day period sinusoidal function, a radial
Mat\'ern-3/2 kernel with a timescale of 20 days was sufficient to ensure the instrumental
signal was removed without inflating the white noise. This is important because the
stellar signal is generally unknown \emph{a priori}, and any attempt to estimate it from the data
will be subject to how well one can discriminate between instrumental and stellar effects.

Finally, we caution that GP regression assumes Gaussian noise. Deviations from Gaussianity,
such as the presence of outliers, can invalidate the assumptions of the method and lead
to poor fits to the data. Although we remove outliers prior to optimizing the GP, users
accessing our de-trended light curves are encouraged to check for Gaussianity with
methods such as an Anderson-Darling test. In \S\ref{sec:methods} we describe our 
iterative procedure for removing outliers and optimizing the GP kernel.

\section{Methods}
\label{sec:methods}
In \S\ref{sec:pld} we outlined the basics of principal component regression
on the fractional pixel fluxes using a GP. In this section, we describe how we apply
this method to \emph{K2} data.

\subsection{Pre-Processing}
\label{sec:pre}
For each star in the \emph{K2} EPIC catalog, we download the target pixel files and
select aperture number 15 from the \texttt{K2SFF} catalog \citep{VJ14, VAN14}, which is derived
by fitting the Kepler pixel response function (PRF) to the stellar image. The \texttt{K2SFF}
data contain twenty such apertures of varying sizes for each star. We find that aperture 15
is typically the best compromise between having enough pixels to generate a good basis
set for PLD while preventing excess contamination from neighboring stars. We note that
when contamination is not an issue, the choice of aperture has little effect on our results.
As in \cite{VJ14}, we then compute the median per-timestamp flux in the pixels of the image 
that lie outside the aperture and subtract it from all pixels
to remove the background signal. In this paper, we consider only the long cadence (LC)
$\emph{K2}$ dataset.

Next, we perform iterative sigma-clipping to mask large outliers. Just
as PLD can artificially inflate the white noise in an attempt to remove stellar
variability (\S\ref{sec:gp}), it can do so in the presence of short timescale features
such as eclipses, transits, flares, or cosmic ray hits. \cite{DEM15} deal with this
by explicitly adding a transit term to the PLD model (Equation~\ref{eq:demingmodel}),
but this requires knowledge of the transit parameters. Since we do not
know \emph{a priori} whether there are transits in a given light curve, we instead 
mask all features in the light curve that are not well modeled by either the PLD
terms or the GP. We then train our model on the masked light curve to compute the
PLD coefficients and use those to de-trend the full, unmasked light curve.

We detect outliers by dividing the light curve into five chunks and de-trending each
with a first order PLD model. We use a 2-day Mat\'ern-3/2
covariance (see \S\ref{sec:gpopt}) with amplitude equal to the median standard deviation
of the flux in all 2-day segments; again, we note that our results are relatively
insensitive to the particular value of these parameters. Next, we perform a
median absolute deviation (MAD) cut at 5$\sigma$ to identify outliers.
We then re-compute the PLD model by masking those outliers and de-trend the full
light curve, identifying a new set of outliers. This process is repeated until no new outliers
are identified. We find that this identifies deep transits, eclipses, and
other outliers that could affect the quality of the PLD fit.

\pagebreak

\subsection{GP Optimization}
\label{sec:gpopt}
\begin{figure*}[t]
  \begin{center}
    \leavevmode
      \psfig{file=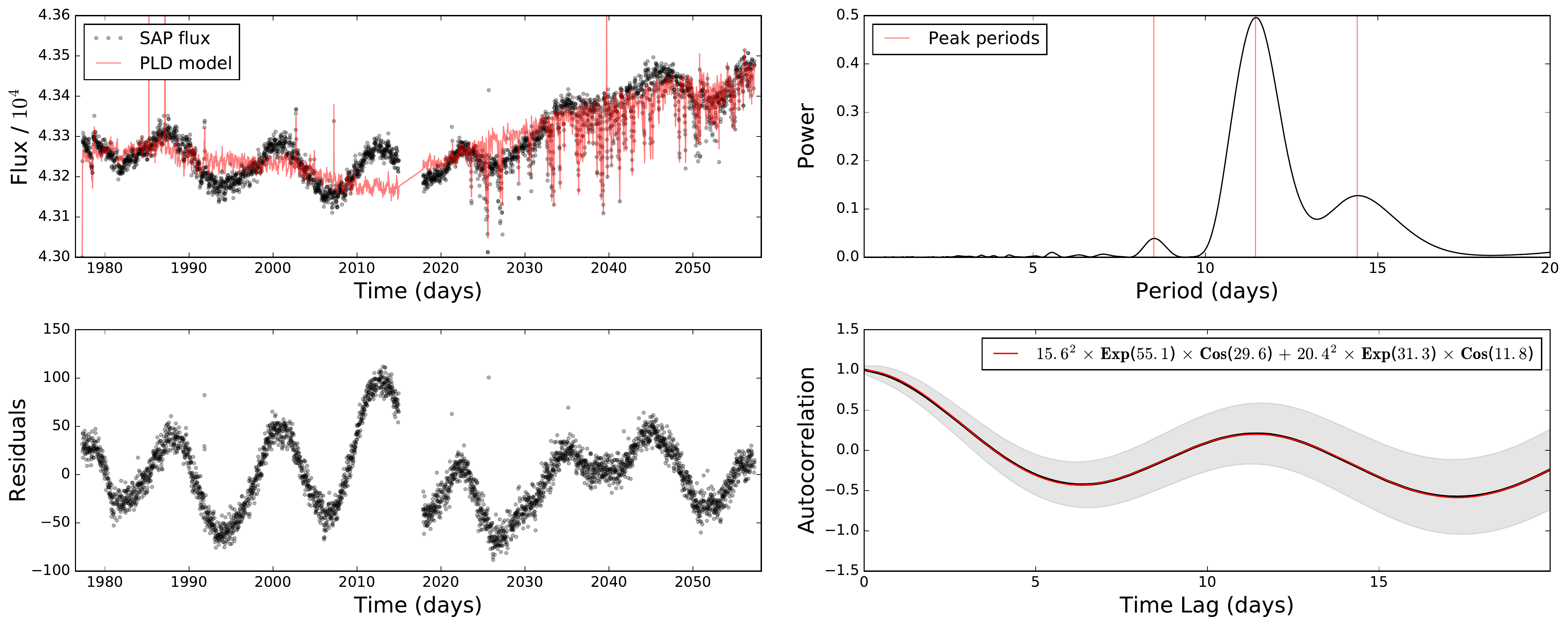,width=\textwidth}
       \caption{GP optimization procedure for EPIC 201497682. In the top left panel we 
                plot the raw SAP flux (black) and a ten chunk, first order PLD fit
                (red); the residuals are shown in the panel below. These are used to
                compute the power spectrum of the stellar signal (top right), and
                its autocorrelation function (bottom right, black curve). Different
                kernels are then fit to the autocorrelation function, and the one
                with the lowest $\chi^2$ value is chosen for the de-trending step
                (red curve). The grey envelope about the autocorrelation curve is
                the ad hoc standard error assumed to compute $\chi^2$.
                }
     \label{fig:acor}
  \end{center}
\end{figure*}

The next step is to compute the covariance matrix $\mathbf{K}$, which we do using
the \texttt{george}\footnote{\texttt{\url{http://dan.iel.fm/george/current/}}} Python package
\citep{GEORGE}. We parametrize 
it as
\begin{align}
\label{eq:covariance}
\mathbf{K}_{ij} = k_w(t_i, t_j) + k_t(t_i, t_j)
\end{align}
where
\begin{align}
\label{eq:whitekernel}
k_w(t_i, t_j) = \sigma^2\delta_{ij}
\end{align}
is a white noise kernel with standard deviation $\sigma$ and $k_t$ is either an 
additive or multiplicative combination
of one or more of an exponential kernel $k_e$, a Mat\'ern-3/2 kernel $k_m$, and
a cosine kernel $k_c$:
\begin{align}
\label{eq:kernels}
k_e(t_i, t_j) &= \alpha_e e^{-\left|t_i - t_j\right|/\tau_e}\nonumber\\
k_m(t_i, t_j) &= \alpha_m \left(1 + \sqrt{3(t_i - t_j)^2}\right) e^{-\sqrt{3(t_i - t_j)^2}/\tau_m}\nonumber\\ 
k_c(t_i, t_j) &= \alpha_c \cos{\left(\frac{2\pi}{P}(t_i - t_j)\right)}.
\end{align}
The choice of kernel depends on the properties of the stellar signal, which we do
not know \emph{a priori}, since it is mixed with the instrumental signal in the
light curve. We therefore adopt an iterative procedure, where we guess at the initial
kernel form and hyperparameters and use it to de-trend the light curve with PLD, thus
obtaining an approximate stellar component. We then train the GP on this stellar
component and use it to run our PLD analysis again. While this can be repeated
multiple times, we find that two iterations are typically enough for most targets. This procedure
is illustrated in Figure~\ref{fig:acor}.

Our initial kernel is a Mat\'ern-3/2 kernel with $\tau_m = 2$ days and amplitude
equal to the median variance in 2-day chunks of the light curve. We split the light curve
into 5-10 roughly equal chunks and de-trend each of them with first order PLD
using Equations~\ref{eq:pcamodel} and \ref{eq:pcacoefficients}. Since we use first order
PLD at this step, we do not perform PCA here. We then compute the 
autocorrelation function of the de-trended data and perform least-squares fits
on different additive and multiplicative combinations of the kernels in Equation~\ref{eq:kernels}, using
the peaks in a Lomb-Scargle periodogram as the initial
guess for the periods in the cosine kernels. We choose the kernel and corresponding
hyperparameters that result in the best fit and compute its overall amplitude
as well as the amplitude of the white noise kernel $k_w$
by maximizing the marginal likelihood $\mathcal{L}$ of the data given the model,
\begin{align}
\label{eq:like}
\log\mathcal{L} = -\frac{1}{2}\mathbf{y}^\top\cdot\mathbf{K^{-1}}\cdot\mathbf{y} - \frac{1}{2}\log\left|\mathbf{K}\right| - \frac{n}{2}\log 2\pi
\end{align}
where $\mathbf{y}$ is the de-trended flux and $n$ is the number of data points 
in $\mathbf{y}$ \citep{RW06}. In principle, one may optimize the parameters of each
of the kernel combinations in this way to obtain the best estimate of the covariance
matrix. However, such a procedure is computationally expensive. After considerable
experimentation, we find that the method outlined above---which takes on the order of one
minute on a single core of a 2.66 GHz machine---is sufficient to ensure PLD removes
only the instrumental component of the noise for most targets.

\subsection{Design Matrix Optimization}
\label{sec:dmopt}
\begin{figure}[h]
  \begin{center}
      \psfig{file=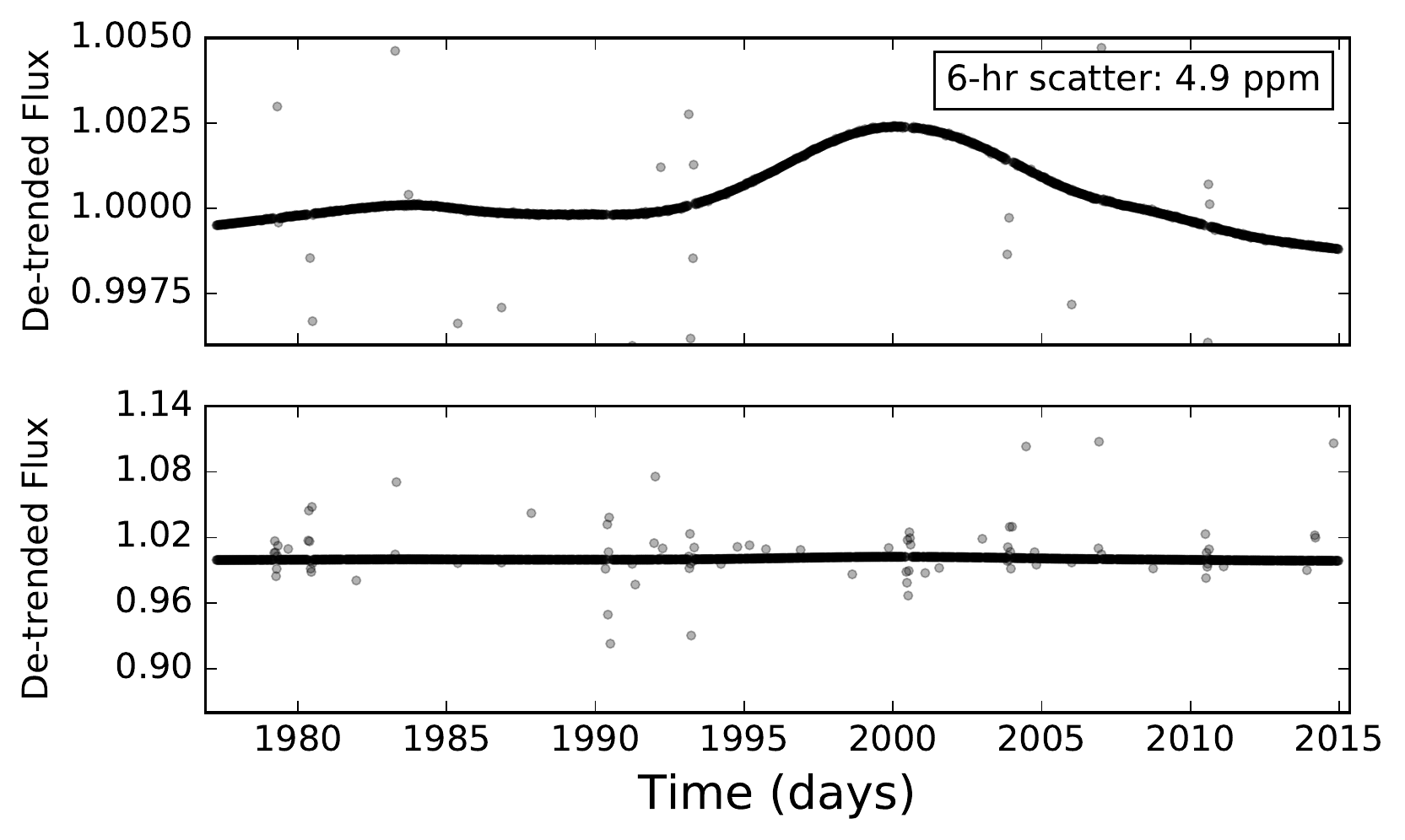,width=3.5in}
       \caption{\emph{Top}: Third order PLD applied to EPIC 201367065, but this time keeping 
                \emph{all} basis vectors. Compare to 
                Figure~\ref{fig:third_order}. While the median scatter improved by a factor
                of about 4, the scatter in the transits (which were masked during the de-trending)
                increased by a factor of several thousand. \emph{Bottom}: The same figure, 
                but zoomed out to show the in-transit scatter.}
     \label{fig:overfitting}
  \end{center}
\end{figure}

The next step in our pipeline is to construct our design matrix 
(Equation~\ref{eq:designmatrix}). In order to do so, we must choose the
number of PLD principal components to use in the regression. This number must be
large enough to capture most of the instrumental variability
but not so large as to lead to overfitting. In fact, one must take care
to prevent PLD (or any other de-trending technique) from fitting out the white noise 
component of the light curve. As the number of basis vectors grows to become very large, 
there begin to exist linear combinations of these vectors that can artificially remove 
white noise from the data, improving the apparent quality of the fit but leading to spurious 
results. This is best illustrated by considering the example in Figure~\ref{fig:third_order}, 
where we kept only 285 basis vectors in the bottom panel. Had we de-trended with all 
8435 components, we would obtain the light curve shown in Figure~\ref{fig:overfitting}.
While the median scatter improved by a factor of $\sim$~4, the in-transit scatter increased
by a factor of several thousand. This is because we masked the transits of K2-3b,
c, and d during the de-trending step (see \S\ref{sec:pre}). The poor
performance of the extrapolated model betrays its terrible predictive power, a clear sign
of overfitting.

\begin{figure}[h]
  \begin{center}
      \psfig{file=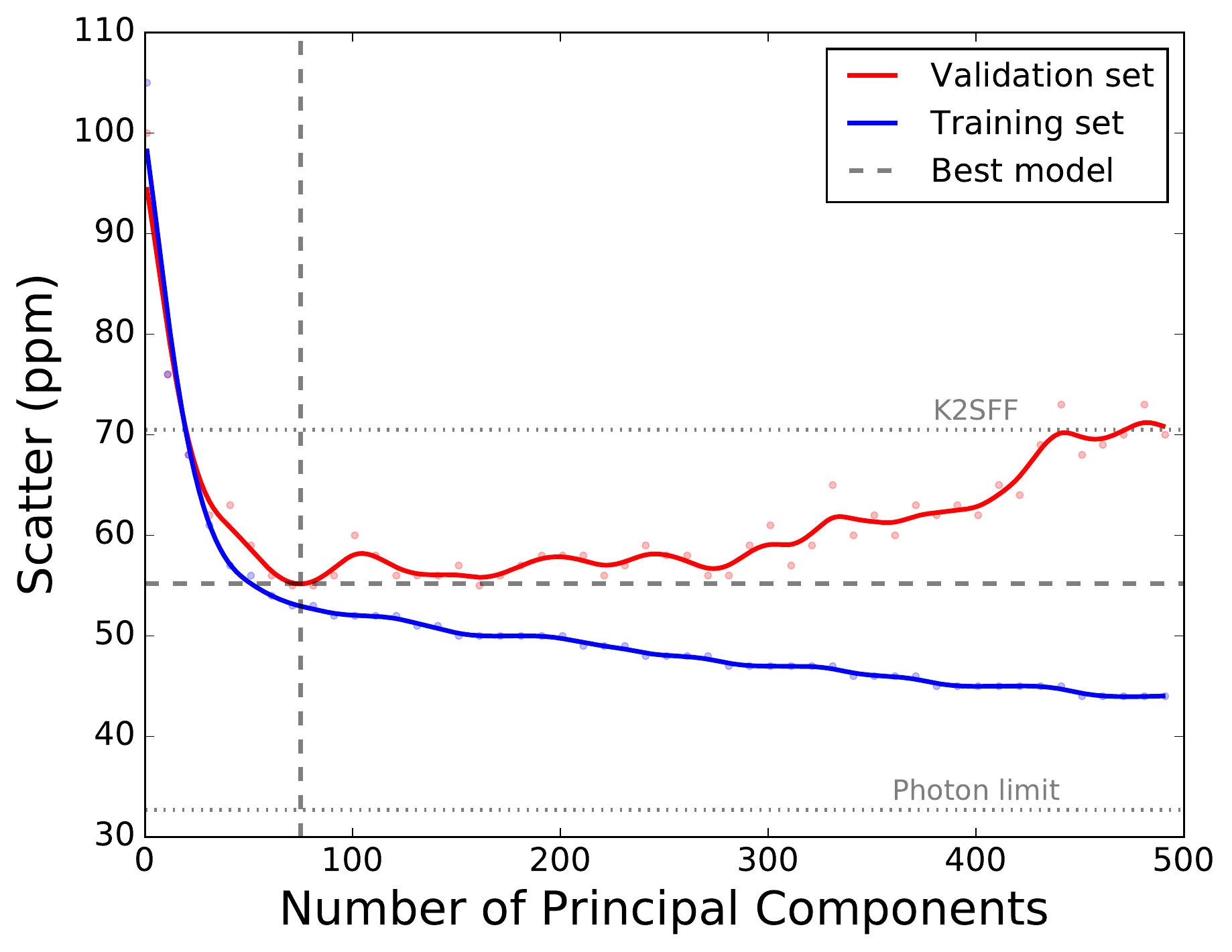,width=3.5in}
       \caption{De-trended light curve precision as a function of the number of principal
                components for EPIC 201497682. The blue dots are the median
                6-hr precision (in ppm) of the unmasked sections of the light curve (the training set); 
                the red dots are the median 
                precision in 6-hr chunks that were masked during the de-trending step
                (the validation set). Solid curves indicate our GP fit to the data points.
                Initially, the scatter decreases in both cases as the 
                number of components is increased. However, above $\sim$ 50 components, while the
                scatter in the training set continues to decrease, the scatter in the validation set (where 
                the model is extrapolated) begins to grow. This is the signature
                of overfitting. We therefore choose 50 principal components for the
                de-trending, yielding a precision of 55 ppm (versus $\sim$ 70 ppm for the
                \texttt{K2SFF} de-trended flux).}
     \label{fig:scatter}
  \end{center}
\end{figure}

In order to limit the flexibility of our PLD model, we implement a simple cross-validation
scheme. We divide the light curve into training sets (large chunks of data that we use to compute the PLD
coefficients) and validation sets (small randomly selected chunks that we de-trend using the coefficients
computed from the training sets). We then compute the 6-hr precision in the validation
sets for a range of design matrix sizes (details below). Since the validation data is never used
to compute the PLD coefficients, the model cannot overfit in the validation set. Instead,
as the number of regressors becomes too large and PLD begins to fit out astrophysical
signals and/or white noise, the scatter in the validation set will grow.
This process is illustrated in Figure~\ref{fig:scatter}, where we plot the scatter in 
the training set (blue) and the scatter in the validation set (red) as a function of
the number of principal components $n_{pc}$.

Initially, as $n_{pc}$ increases, the scatter in both the training and validation data
decreases. As the number of components increases further, the training precision continues to improve,
eventually surpassing the photon noise limit for $n_{pc} \gtrsim 1000$. This is obviously unphysical and a
clear sign of overfitting. The scatter in the validation set, on the other hand, begins to increase steadily above $n_{pc} \sim 50$,
indicating that the predictive power of the model worsens as the number of principal components is increased.
The minimum in the red curve, which occurs for $n_{pc} \sim 80$, is the best we can do; above that point, PLD
is likely to begin fitting out white noise.

We employ this cross-validation method for each EPIC target. In practice, we compute the 6-hr precision
(see \S\ref{sec:precision}) in the validation sets fifty times for each value of $n_{pc}$ and take the median. 
Each validation set is a group of 10 non-overlapping chunks that are masked when computing the PLD coefficients; each
chunk is chosen randomly from the set of all contiguous 13-cadence segments of the light curve. 
Our grid in $n_{pc}$ typically ranges from 1 to 250 principal components, with a spacing of ten components.
After computing the median scatter for each value of $n_{pc}$, we smooth the curves with a GP and choose the 
value of $n_{pc}$ that minimizes the scatter in the validation set.

Finally, for campaign 1 we found it necessary to split all light curves into two separate
timeseries, with a breakpoint at $\mathrm{BJD} - 2454833 = 2017$, a mid-campaign data gap. 
Inspection of the top
left panel of Figure~\ref{fig:acor} reveals a qualitative change in the instrumental
systematics in the second half of the campaign, likely due to a reversal in the direction
of the spacecraft roll \citep[see, e.g.,][]{AIG16}. While similar reversals are also present 
in other campaigns, campaign 1 is the only one in which we obtain a de-trending improvement 
significant enough to justify the breakpoint. 


\subsection{De-trending}
After optimizing the GP and choosing the order of the PLD, the number of principal components, and the 
number of light curve subdivisions, we de-trend each EPIC light curve by subtracting the
model given by
Equations~\ref{eq:pcamodel} and \ref{eq:pcacoefficients}. We then add the median value of the
SAP flux to the result to obtain our final, de-trended light curves.

\pagebreak

\section{Results}
\label{sec:results}
\subsection{Transit Injections}
\label{sec:injections}
Since the PLD basis vectors are obtained from the \emph{fractional} pixel fluxes,
they do not in principle contain any astrophysical signals. De-trending with PLD
should therefore preserve the transit shape and depth, a fact that is confirmed
for the transits of several hot Jupiters observed with \emph{Spitzer} \citep{DEM15}.
However, as we showed in \S\ref{sec:dmopt}, PLD can increase the
scatter of regions of the data that are not used to compute the coefficients
(such as transits that are flagged as outliers) when too many principal components 
are present (see Figure~\ref{fig:overfitting}). If a transit is missed in the
outlier detection step, the transit shape may also be affected,
since PLD can exploit linear combinations of the white noise component
to decrease the transit depth and therefore improve the likelihood of the fit.

\begin{figure}[h]
  \begin{center}
      \psfig{file=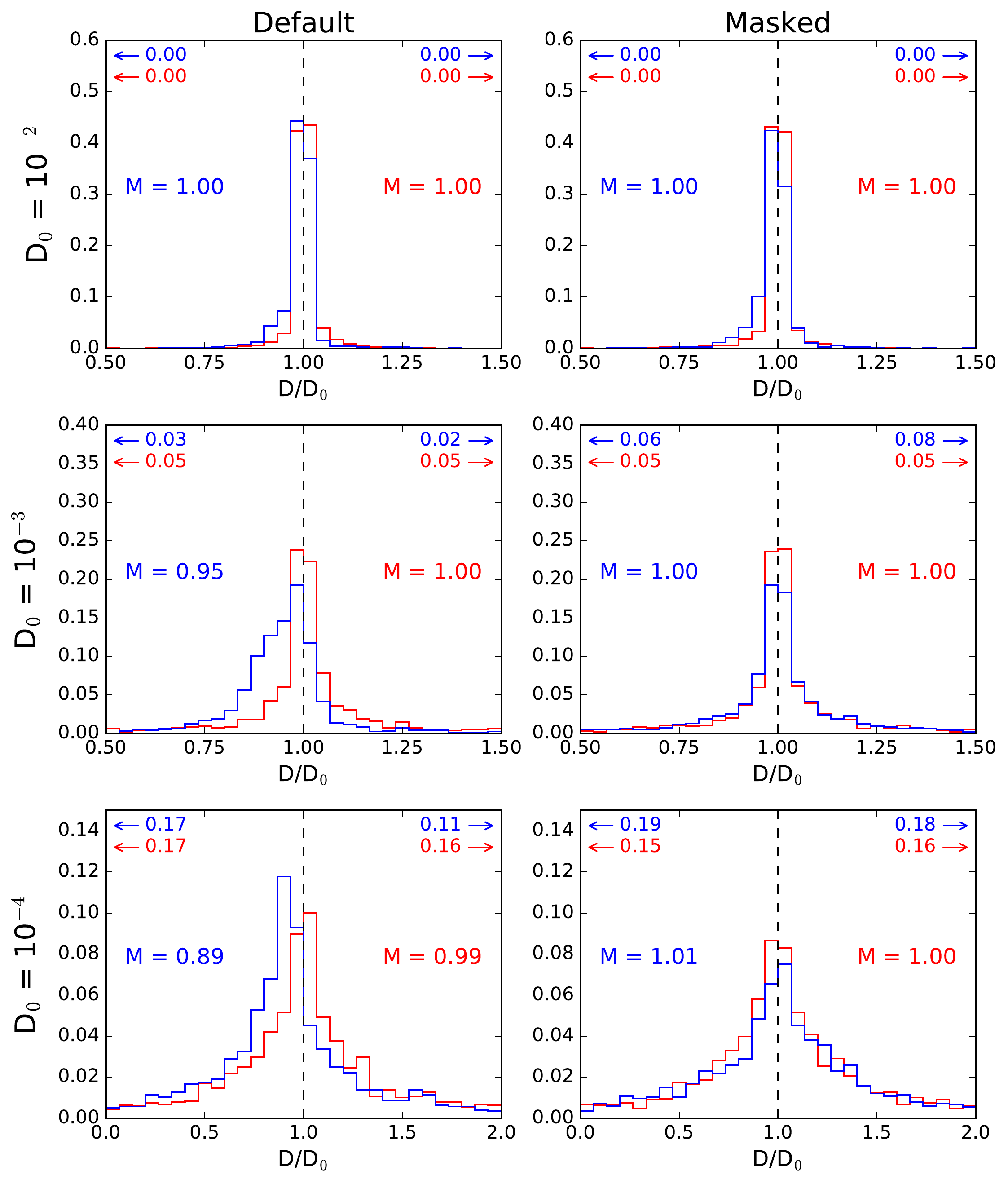,width=3.5in}
       \caption{Transit injection results. Each panel shows the fraction of transits
                recovered with a certain depth ratio $D/D_0$ (recovered depth divided
                by true depth). Blue histograms correspond to the actual injection
                and recovery process performed with our pipeline; red histograms
                correspond to transits injected directly into the de-trended
                light curves and are shown for comparison. The values to the left
                and right of each histogram are the median $D/D_0$ for our pipeline
                and for the control run, respectively. The smaller values at the
                top indicate the fraction of transits recovered with depths lower and
                higher than the bounds of the plots. Finally, the two columns distinguish
                between the default runs (left) and runs
                where the transits were explicitly masked (right); the three rows correspond to different
                injected depths: $10^{-2}$, $10^{-3}$, and $10^{-4}$. PLD preserves
                transit depths if the transits are properly masked; otherwise, a small bias
                toward smaller depths is introduced.}
     \label{fig:injections}
  \end{center}
\end{figure}

In order to test whether our pipeline is robust against these effects, we ran a set
of transit injection and recovery tests on a sample of EPIC
targets that do not contain known transit events. We randomly selected 200 campaigns 0-6 stars 
from each magnitude bin in the range $8 \le K_p \le 18$ ($K_p$ is the Kepler bandpass magnitude)
for a total sample size of 2000 stars per run. We then injected synthetic transits of varying depths into the
light curve by multiplying each pixel by a transit model generated by the 
\texttt{pysyzygy}\footnote{\texttt{\url{https://github.com/rodluger/pysyzygy}}}
package, which calculates fast limb-darkened light curves based on the \cite{MA02} transit model. 
We randomly chose orbital periods in the range 3-10 days and fixed
the transit duration at 2.5 hours, assuming zero eccentricity
and quadratic limb darkening parameters $a = 0.45$ and $b = 0.30$. We then
ran our pipeline to de-trend the light curves.

Performing a full transit search is outside the scope of this paper. Since our 
goal is to determine whether or not PLD can bias transit depths, we fix all the
parameters except for the depth at their true values and recover the transit
depth by linear regression, simultaneously fitting the baseline flux in the
vicinity of the transit with a third-order polynomial. Our results are shown
in Figure~\ref{fig:injections}.

Each panel shows two histograms of the recovered depth as a fraction of the true
depth, $D/D_0$. In blue, we plot the recovery results after de-trending with our pipeline.
As a control, we also injected transits directly into the \emph{de-trended} data; we 
plot the corresponding distributions in red. Provided \texttt{EVEREST} does not affect
transit depths or increase in-transit scatter, these two sets of distributions should 
be similar. The numbers at the top left and right of each
panel respectively indicate the fraction of recovered depths below and above the limits 
of the plot. The median values $M$ of each distribution are also shown.

Each row in the figure corresponds to a different injected depth: 
$10^{-2}$, $10^{-3}$, and $10^{-4}$, ranging from a typical 
hot Jupiter to a roughly Earth-sized planet. The left column corresponds to the default
runs of our pipeline; the right column corresponds to runs in which the transits were
explicitly masked (more details on this below).

In the top left panel ($D_0 = 10^{-2}$), both distributions have medians equal to 1.00, corresponding
to an unbiased recovery of the transit depth. Moreover, the spread is similar in both
distributions, indicating that \texttt{EVEREST} does not introduce significant in-transit
scatter. In the next two panels ($D_0 = 10^{-3}$ and $10^{-4}$), however, a bias toward
smaller transit depth is visible; the \texttt{EVEREST} de-trending causes transits
of these small planets to be shallower by $\sim 5$ and $\sim 10\%$, respectively. This
is because our sigma-clipping outlier detection technique (\S\ref{sec:pre}) is not effective 
at finding shallow transits, and thus these transits are not masked during the de-trending
step. As we mentioned above, since we have no term in our design
matrix (Equation~\ref{eq:designmatrix}) to explicitly model transits, the PLD model
picks up the slack and partially fits out these features by inflating the white noise,
slightly improving the quality of the overall fit. This is similar to the example in 
Figure~\ref{fig:sinusoid}, where PLD inflated the white noise to remove an astrophysical
signal. 

When the transits are properly masked (right column of the figure), both the bias
and the increased scatter in the recovered depths disappear for transits of all
depths. This shows that our cross-validation technique (\S\ref{sec:dmopt}) is
correctly preventing overfitting and enforcing the high predictive power of the model
in the masked regions of the dataset.

In order to remove the small bias introduced in the presence of shallow, unmasked
transits, we strongly encourage those making use of our light curves for transiting exoplanet
characterization to run \texttt{EVEREST} while explicitly masking these transits. 
This can be done by specifying the transit times and durations when calling the \texttt{EVEREST}
Python module; the code takes only a few seconds to run. The same applies to those using our light
curves for transiting planet searches. Once features of interest are detected, one
should run \texttt{EVEREST} again with those features masked to obtain unbiased
estimates of the transit parameters.

It is important to note, however, that some transits with very low signal-to-noise could 
be completely fit out during the de-trending step, preventing their detection in 
the first place. As \cite{DFM15} point out, this is an inevitable
consequence of the de-trend-then-search method. It is \emph{always} best to use
a model that captures all the features of the data, allowing one to solve for 
instrumental noise, stellar variability, and transits \emph{simultaneously}. To this end
\cite{DFM15} explicitly include a transit model in their design matrix, solving
Equation~\ref{eq:pcacoefficients} over a fine grid of periods and transit epochs. This
eliminates the de-trending step in favor of a combined de-trending/planet
searching step, which effectively circumvents the overfitting problems inherent to
least-squares de-trending techniques. However, such an approach is very computationally
expensive, given that each light curve must be processed once for every combination
of transit parameters. We therefore defer this procedure to a future paper.

\begin{figure}[h]
  \begin{center}
      \psfig{file=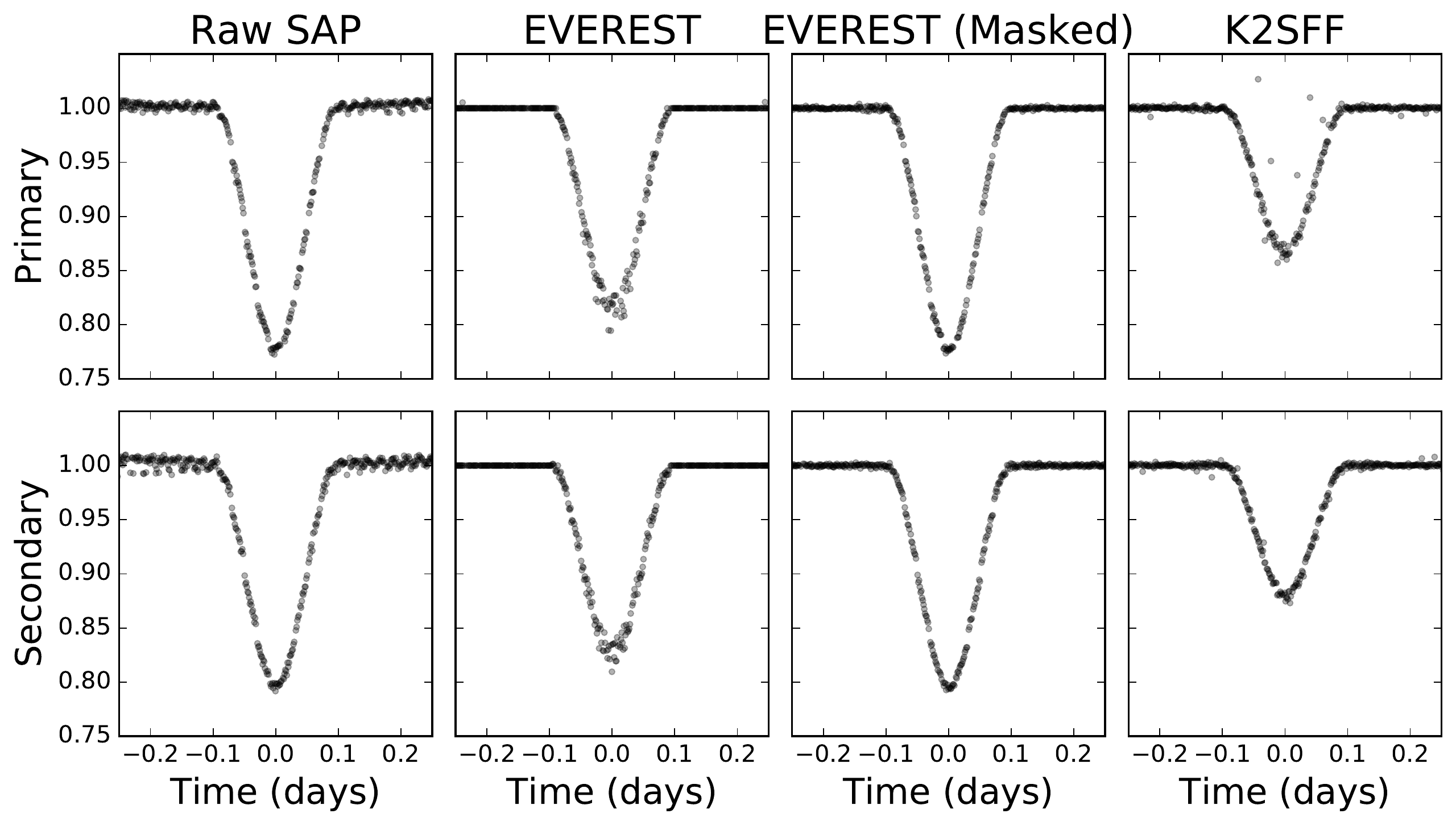,width=3.25in}
       \caption{Primary and secondary eclipses of EPIC 202072563, folded
                on a period of 2.1237 days. From left to right, the columns
                show the raw SAP flux, the \texttt{EVEREST} flux, the \texttt{EVEREST}
                flux obtained when explicitly masking the eclipses, and
                the \texttt{K2SFF} flux. Masking
                is critical to preserving the transit depth during de-trending
                for any pipeline that does not explicitly include a transit or
                eclipse model.}
     \label{fig:202072563}
  \end{center}
\end{figure}

We note, finally, that other \emph{K2} pipelines are subject to similar overfitting
problems. Consider the example in Figure~\ref{fig:202072563}, which shows 
the (folded and normalized) primary and secondary eclipses of the eclipsing binary EPIC 202072563. 
Four datasets are presented: the raw SAP flux (left), the default \texttt{EVEREST} light curve, 
the masked \texttt{EVEREST} light curve, and the \texttt{K2SFF} light curve. 
The \texttt{EVEREST} and \texttt{K2SFF} fluxes have been smoothed with a GP to 
remove astrophysical variability. It is clear from the second column that our outlier detection technique failed to 
mask all points during the eclipse, leading to a $\sim 20\%$ decrease in the eclipse
depths. When properly masking the eclipses, however, the depth is correctly recovered 
(third column). In the last column, we see that the \texttt{K2SFF} depths are also
shallower, but by nearly $50\%$. Proper modeling or masking of these features would be necessary to
preserve their depths during the \texttt{K2SFF} de-trending step.

\subsection{Limitations}
\label{sec:limitations}
We find that there are two particular situations in which PLD is likely to fail: saturated stars
and crowded apertures. These limitations are inherent to the method itself and not
specific to \emph{K2} data.

\subsubsection{Saturated Stars}
\label{sec:saturated}
\begin{figure}[h]
  \begin{center}
      \psfig{file=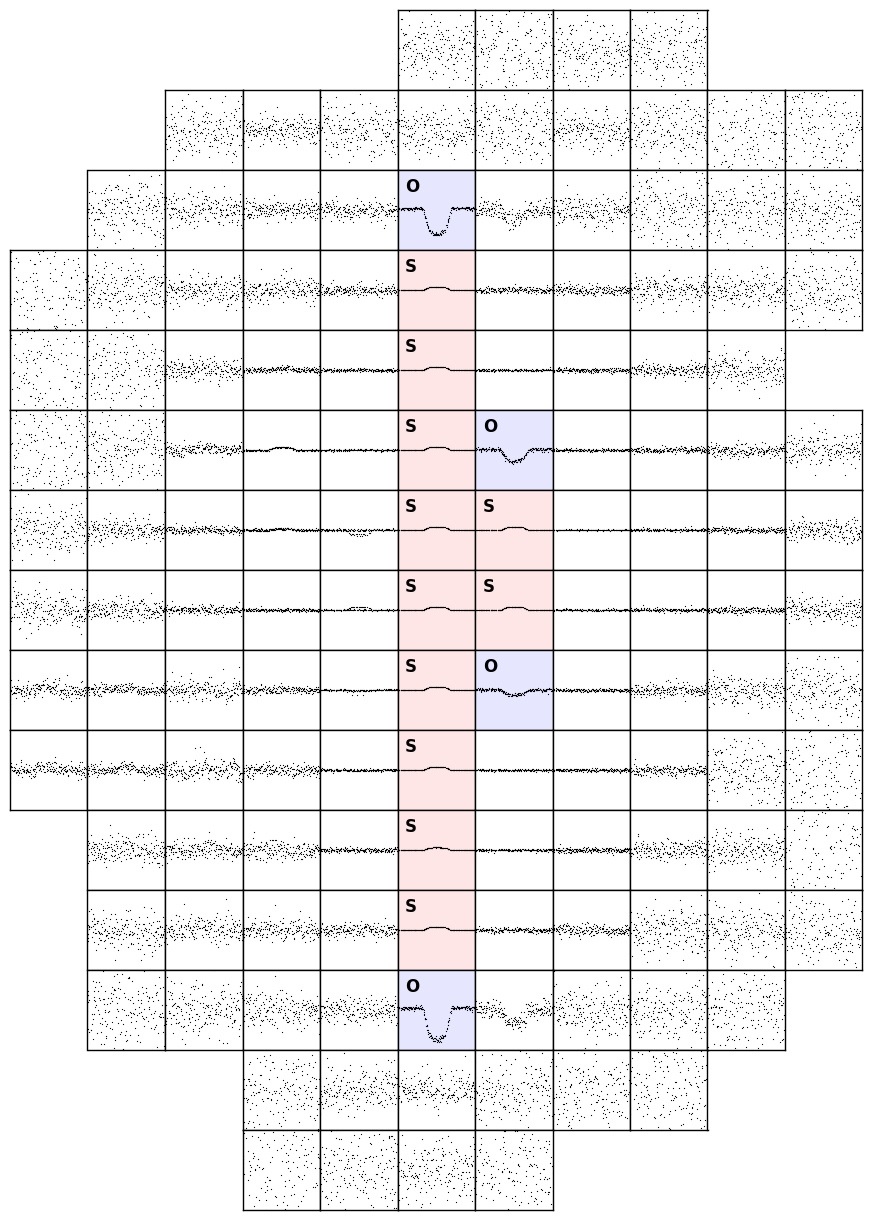,width=3.25in}
       \caption{Fractional pixel fluxes $p_{il} / \sum_k{p_{ik}}$
                for quarter 3 of Kepler-3, a $K_p = 9.2$ hot-Jupiter host
                observed by the original \emph{Kepler} mission. The panels
                are arranged according to the positions of the pixels on
                the detector, and the data is smoothed and folded on the orbital period
                of Kepler-3b. Saturated pixels are highlighted in red and are labeled
                with an \textbf{S};
                overflow pixels are highlighted in blue and labeled with an \textbf{O}. PLD fails for this
                system because the transit signal is present in several of
                the basis vectors.}
     \label{fig:saturation}
  \end{center}
\end{figure}

The \emph{Kepler} detectors begin to saturate for stars with $K_p \lesssim 11.5$,
leading to flux bleeding along the pixel columns \citep{GIL10}. Since the total
charge is well conserved for stars dimmer than $K_p \approx 7$, this is not an issue for
aperture photometry; transit depths and shapes are preserved in the SAP flux.
However, the basic assumption of PLD---that the fractional pixel fluxes 
$p_{il} / \sum_k{p_{ik}}$ (see Equation~\ref{eq:pldmodel}) contain no astrophysical 
information---breaks down for these stars. This occurs because saturated pixels 
contain virtually no transit signal, as both the in-transit flux and the out-of-transit
flux are above the saturation level, resulting in a relatively featureless
timeseries. Since the total flux is conserved, the transit signal 
overflows into the adjacent pixels, traveling along the column until it reaches
an unsaturated pixel. In these ``overflow'' pixels, the fractional transit
depth is much larger than the true depth, since it contains the total transit signal
from all of the saturated pixels in that column. Consequently, the normalization
$p_{il} / \sum_k{p_{ik}}$ will only \emph{partially} remove the transit in the
``overflow'' pixels. Conversely, it will over-correct the saturated pixel fluxes,
leading to an inverted transit shape in the PLD basis vectors corresponding to 
those pixels. This is illustrated in Figure~\ref{fig:saturation}, which shows the 
\emph{fractional} pixel fluxes as a function of their position on the detector for the saturated
hot-Jupiter host Kepler-3b. We again choose a \emph{Kepler} target for illustrative
purposes, though the idea applies equally to \emph{K2}. Saturated pixels are indicated in red; ``overflow'' pixels
are indicated in blue; both groups of pixels contain the transit signal. Unsurprisingly, 
PLD fails to properly de-trend this target, removing most of the transit signal
along with the instrumental noise.

\begin{figure}[h]
  \begin{center}
      \psfig{file=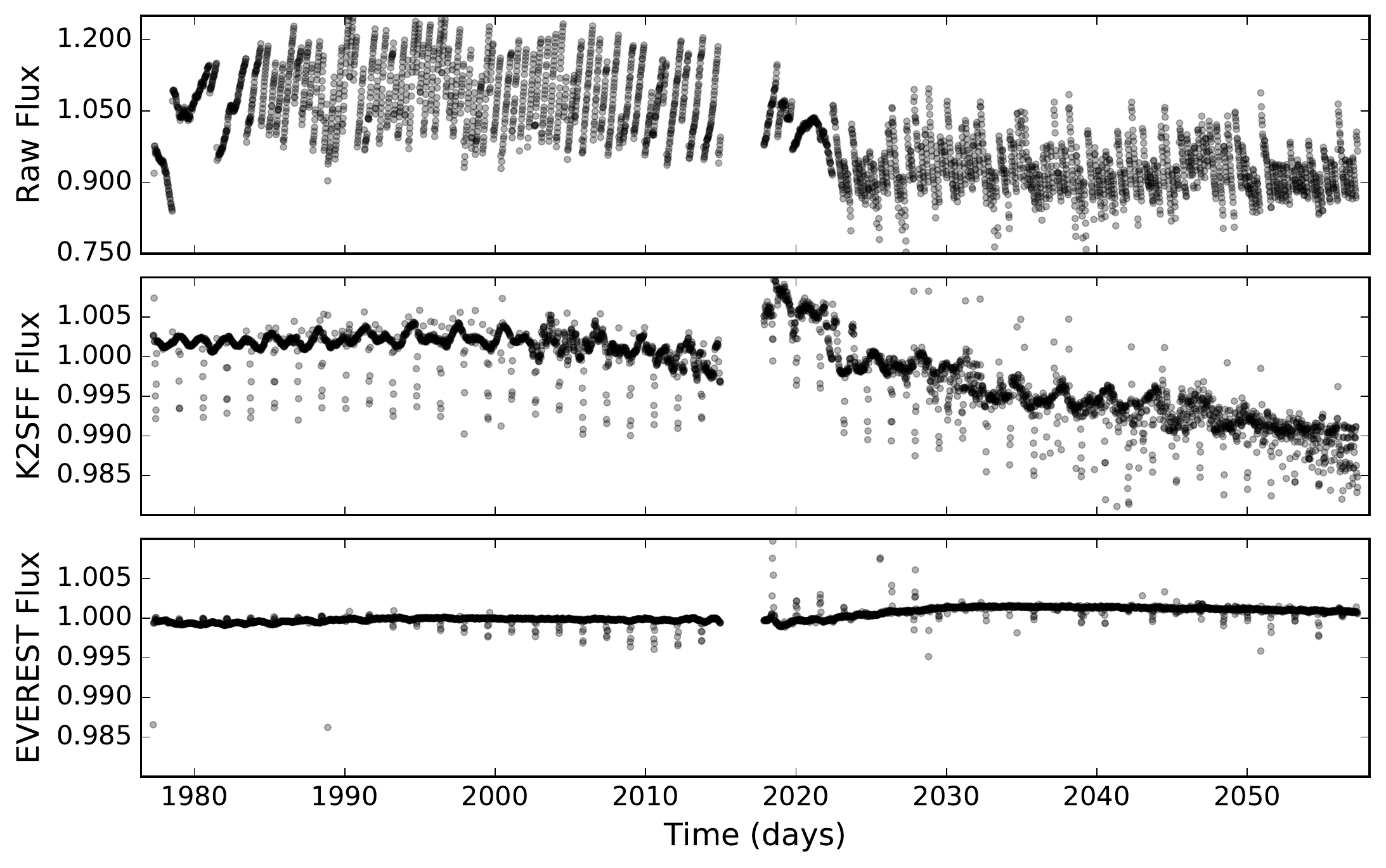,width=3.5in}
       \caption{EPIC 201270464, a $K_p = 9.4$ saturated eclipsing binary. Plotted here
                is the raw flux (top), the \texttt{K2SFF} flux (center), and the \texttt{EVEREST}
                flux (bottom). PLD washes out the stellar variability along with most of the eclipses
                for some saturated stars.}
     \label{fig:201270464}
  \end{center}
\end{figure}

In Figure~\ref{fig:201270464}, we show an example of a saturated light curve of an eclipsing
binary processed with \texttt{K2SFF} and \texttt{EVEREST}. The \texttt{K2SFF} pipeline removes
a significant amount of the instrumental noise while preserving most of the astrophysical signal.
\texttt{EVEREST}, on the other hand, fits out both the stellar variability and the eclipses,
leading to spurious low scatter.

In principle, the saturation effects may be mitigated by collapsing each saturated column
into a single timeseries; since charge is conserved within individual columns, this would enforce
the correct transit depth in those ``superpixels,'' which could then be used as basis
vectors for PLD. However, we leave this procedure to future work. At present, we flag
saturated stars in our database that are at risk of PLD overfitting. As a general rule, 
while saturation may occur for stars as faint as $K_p = 11.5$, we find that PLD preserves transit
depths for most stars dimmer than $K_p = 11$.

\subsubsection{Crowded Apertures}
\label{sec:crowded}
The other situation in which PLD performs poorly is in apertures containing significant
contamination from other stars. In simple aperture photometry, the transit depth
can be decreased in the presence of another star within the aperture. To correct for 
this, one can simply scale the depth by the (inverse of the) crowding metric, the ratio of 
the flux due to the planet host to the total flux in the aperture. However, this is not
the case for data de-trended with PLD, for reasons we describe below.

In the case of a single star, the pixel flux $p_{il}$ in the $l^\mathrm{th}$ pixel at the $i^\mathrm{th}$ time
is the product of the (position-dependent) stellar signal $a_{il}$ and the (position-independent) transit signal $\tau_{i}$,
such that the fractional pixel flux is
\begin{align}
\label{eq:crowdingonestar}
\frac{p_{il}}{\sum\limits_{k}p_{ik}} = \frac{a_{il}\tau_i}{\sum\limits_{k}a_{ik}\tau_i} = \frac{a_{il}}{\sum\limits_{k}a_{ik}},
\end{align}
which, as we have argued before, contains no transit signal. In the case of two stars, $a_{il}$ and $b_{il}$, 
the first of which contains a transit, we may write the fractional flux as
\begin{align}
\label{eq:crowdingtwostars}
\frac{p_{il}}{\sum\limits_{k}p_{ik}} = \frac{a_{il}\tau_i + b_{il}}{\sum\limits_{k}a_{ik}\tau_i + b_{ik}}.
\end{align}
Note that in this case the transit signal does not cancel, and the PLD basis vectors will contain some amount of
transit information, leading to possible overfitting as in the case of a saturated aperture. For a 
dim contaminant star, $b_{il} \ll a_{il}$, we may write
\begin{align}
\label{eq:crowdingtwostarsapprox}
\frac{p_{il}}{\sum\limits_{k}p_{ik}} \approx \frac{a_{il}}{\sum\limits_{k}a_{ik}}\left(1 + \frac{\Delta}{\tau_i}\right)
\end{align}
where\begin{align}
\label{eq:deltacrowding}
\Delta \equiv \left(\frac{b_{il}}{a_{il}} - \frac{\sum\limits_{k}b_{ik}}{\sum\limits_{k}a_{ik}}\right)
\end{align}
is a measure of the difference between the crowding in a given pixel and the crowding
in the entire aperture. In general, the larger the value of $\Delta$, the more power
PLD will have to fit out the transit signal. This is the case for \emph{bright contaminant
sources near the edge of the aperture}, for which the value of 
$b_{il}/a_{il}$ varies greatly across the aperture. Interestingly, for contaminant
sources co-located with the transiting planet host (as in the case of binary stars
or stars that are aligned but unresolved), the quantity $b_{il}/a_{il}$ is constant
across the detector and $\Delta = 0$, leading to a PLD basis set that does not contain
transit information. This can also be shown from the exact expression 
(Equation~\ref{eq:crowdingtwostars}). Crowding is therefore only a concern when
there is a bright contaminant sufficiently separated from the transiting planet host.
In practice, we find that PLD begins to overfit for contaminants that are 
separated by more than one pixel from the target and are either brighter than
or within $\sim 2$ magnitudes of the target.

It is possible to correct for the effects of crowding if the quantity $b_{il}/a_{il}$
is known, even if just approximately. However, this requires careful modeling of the
stellar PSF and is beyond the scope of this paper. In our database, we flag sources
that may suffer from overfitting due to crowded fields.

\subsection{Photometric Precision}
\label{sec:precision}
The formal photometric noise metric developed by the \emph{Kepler} team is the
Combined Differential Photometric Precision (CDPP) \citep{CHR12}, which is computed
by the \emph{Kepler} pipeline on transit-like timescales of 3, 6 and 12 hours.
The CDPP evaluated for a given duration is defined so that it is equivalent to the 
depth of a transit of that duration that would yield a SNR of 1. In this section we adopt a proxy 
of the 6-hr CDPP that is easier to calculate than
the formal metric defined in \cite{CHR12}. Our approach is very similar to the approaches
of \cite{GIL11}, \cite{VJ14} and \cite{AIG16}. In order to prevent correlated stellar
noise from inflating the white noise calculation, we apply a 2-day, quadratic Savitsky-Golay 
\citep{SG64} high-pass filter to the de-trended flux, clipping
outliers at 5$\sigma$. We then compute the standard deviation of all contiguous 13-cadence
chunks of data, take the median, and divide by $\sqrt{13}$ as in \cite{VJ14} to obtain
the approximate 6-hr photometric precision of the data, which we henceforth refer to
as the CDPP.

\label{sec:precision}
\begin{figure}[h]
  \begin{center}
      \psfig{file=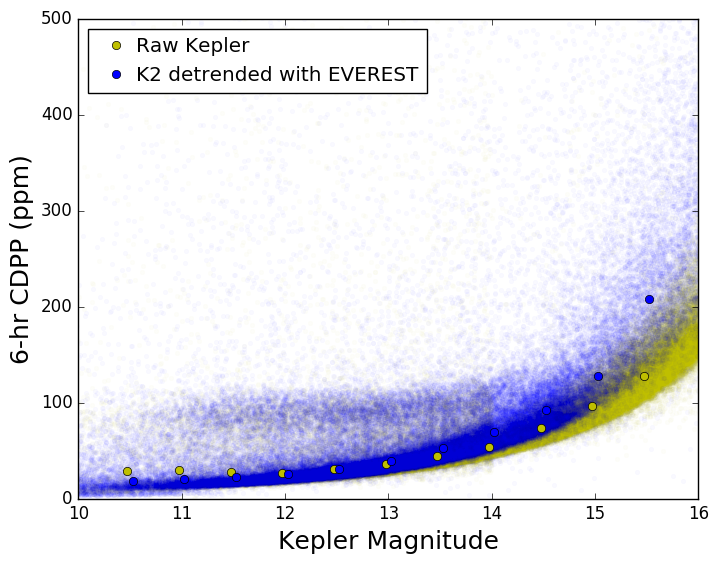,width=3.5in}
       \caption{6-hr photometric precision as a function of Kepler magnitude $K_p$
                for all stars observed by \emph{Kepler} 
                (yellow dots) and for all unsaturated, non-crowded \emph{K2} targets
                in campaigns 0-7 de-trended with \texttt{EVEREST}
                (blue dots). The median values are indicated
                for 0.5 magnitude-wide bins with filled circles.
                Our pipeline recovers the original \emph{Kepler} precision for stars
                brighter than $K_p \approx 13$.}
     \label{fig:precision1}
  \end{center}
\end{figure}
\begin{figure}[h]
  \begin{center}
      \psfig{file=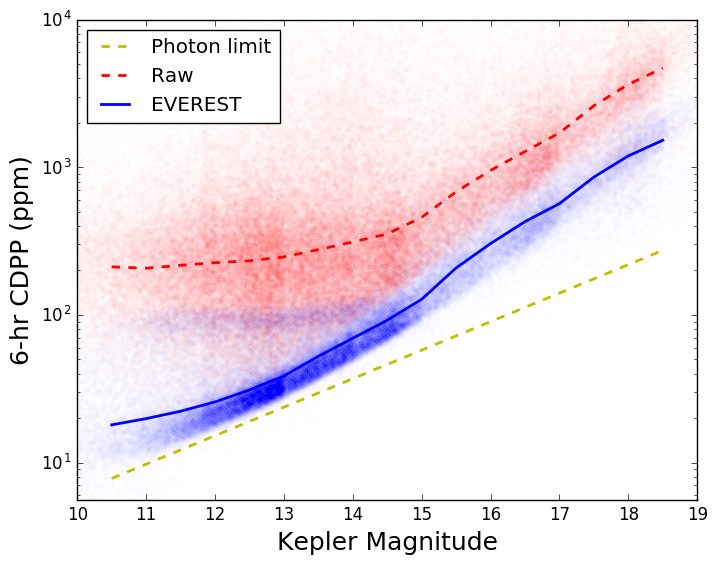,width=3.5in}
       \caption{A comparison of the raw \emph{K2} 6-hr precision (red dots) and the \texttt{EVEREST} 
       precision (blue dots) as a function of $K_p$. The lines
       indicate the median in 0.5 magnitude-wide bins. We also plot the approximate photon
       limit (dashed yellow line) for reference. \texttt{EVEREST} leads to an order-of-magnitude
       improvement in the CDPP for the brightest stars.}
     \label{fig:precision2}
  \end{center}
\end{figure}

In Figure~\ref{fig:precision1} we plot the CDPP of our de-trended fluxes for campaign 0-7 
stars, excluding saturated targets (stars whose brightest pixel's median flux $> 1.6\times 10^5 \mathrm{e/s}$) 
and stars in highly crowded apertures (stars with $\Delta K_p < 5$ neighbors
inside the aperture or brighter neighbors within 2 pixels of the edge of the aperture),
since PLD is likely to lead to artificially low CDPP in those cases. \texttt{EVEREST}
CDPP values are plotted as blue dots. For comparison, in yellow we plot the CDPP calculated for the raw
SAP fluxes of all stars observed by the original \emph{Kepler} mission. 
\texttt{EVEREST} recovers the raw \emph{Kepler} CDPP for most $K_p \lesssim 13$ stars
and yields light curves with CDPP within about 1.5 times that of \emph{Kepler}
for $13 \lesssim K_p \lesssim 15$. The sparser clump of stars above the bulk group between
$K_p = 11$ and $14$ are likely giants, whose short-timescale pulsations are not efficiently
captured by the high-pass filter and thus appear to be more noisy; this clump is also
present in the \emph{Kepler} distribution.

In Figure~\ref{fig:precision2} we again plot the CDPP as a function of $K_p$, but this time
on a logarithmic scale, comparing the \texttt{EVEREST} values (blue dots) to the raw \emph{K2}
CDPP (red dots) and the approximate 6-hr photon limit (dashed yellow line), which we calculate as
\begin{align}
\label{eq:phot}
\mathrm{CDPP}_{phot} = \frac{10^6}{\sqrt{21600\times\bar{F}}}
\end{align}
where $\bar{F}$ is the average SAP flux in $\mathrm{e/s}$ at a particular value of $K_p$. The dashed red
and solid blue lines indicate
the median CDPP in 0.5 magnitude-wide bins in the raw and de-trended light curves, respectively.
\texttt{EVEREST} leads to nearly an order-of-magnitude
improvement in the CDPP for the brightest targets, approaching the photon limit for 
$K_p \lesssim 13$ dwarfs.

\subsection{Comparison to Other Pipelines}
\label{sec:comparison}
\begin{figure}[h]
  \begin{center}
      \psfig{file=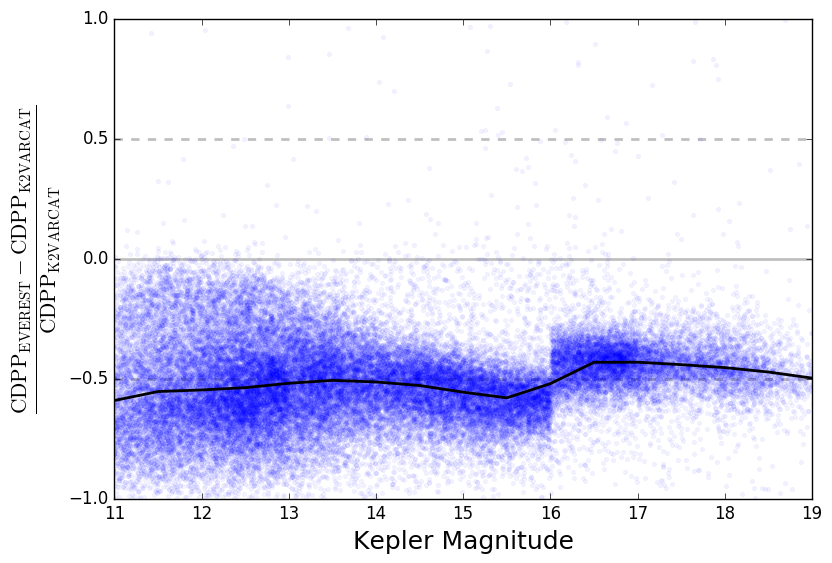,width=3.5in}
       \caption{Relative 6-hr CDPP difference between \texttt{EVEREST}
                and \texttt{K2VARCAT} light curves for campaigns 0-7. Blue dots
                show differences for individual stars, while the black line
                indicates the median in 0.5 magnitude-wide bins. Negative
                values indicate higher precision in the \texttt{EVEREST}
                light curves; compare to Figure~10 in \cite{AIG16}. On average, \texttt{EVEREST} yields
                light curves with half the scatter for all Kepler magnitudes $K_p > 11$.}
     \label{fig:comparison_k2varcat}
  \end{center}
\end{figure}

\begin{figure}[h]
  \begin{center}
      \psfig{file=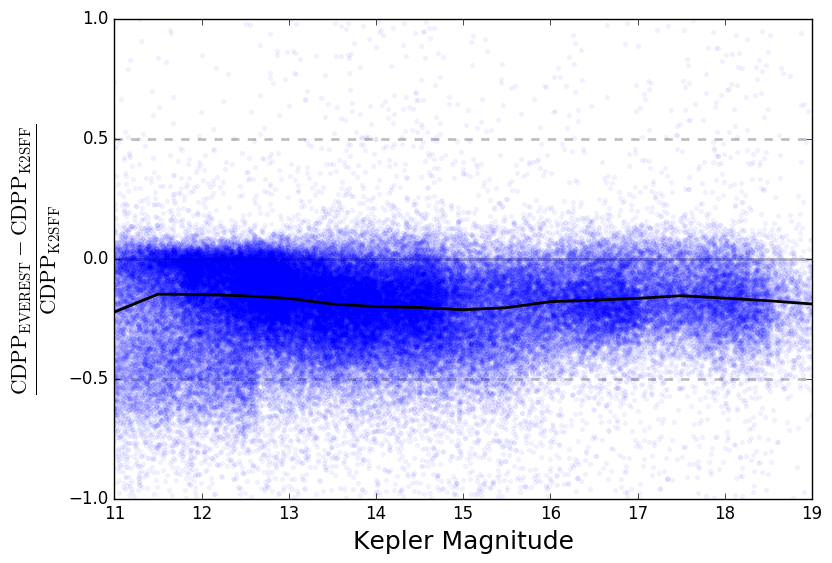,width=3.5in}
       \caption{Same as Figure~\ref{fig:comparison_k2varcat}, but comparing 
                \texttt{EVEREST} to \texttt{K2SFF}. Once again, negative values
                correspond to higher precision in the \texttt{EVEREST} light curves.
                Our pipeline yields higher
                precision light curves for most \emph{K2} stars and does
                better on average for all Kepler magnitudes $K_p > 11$.}
     \label{fig:comparison_k2sff}
  \end{center}
\end{figure}

\begin{figure*}[t]
  \begin{center}
      \leavevmode
      \psfig{file=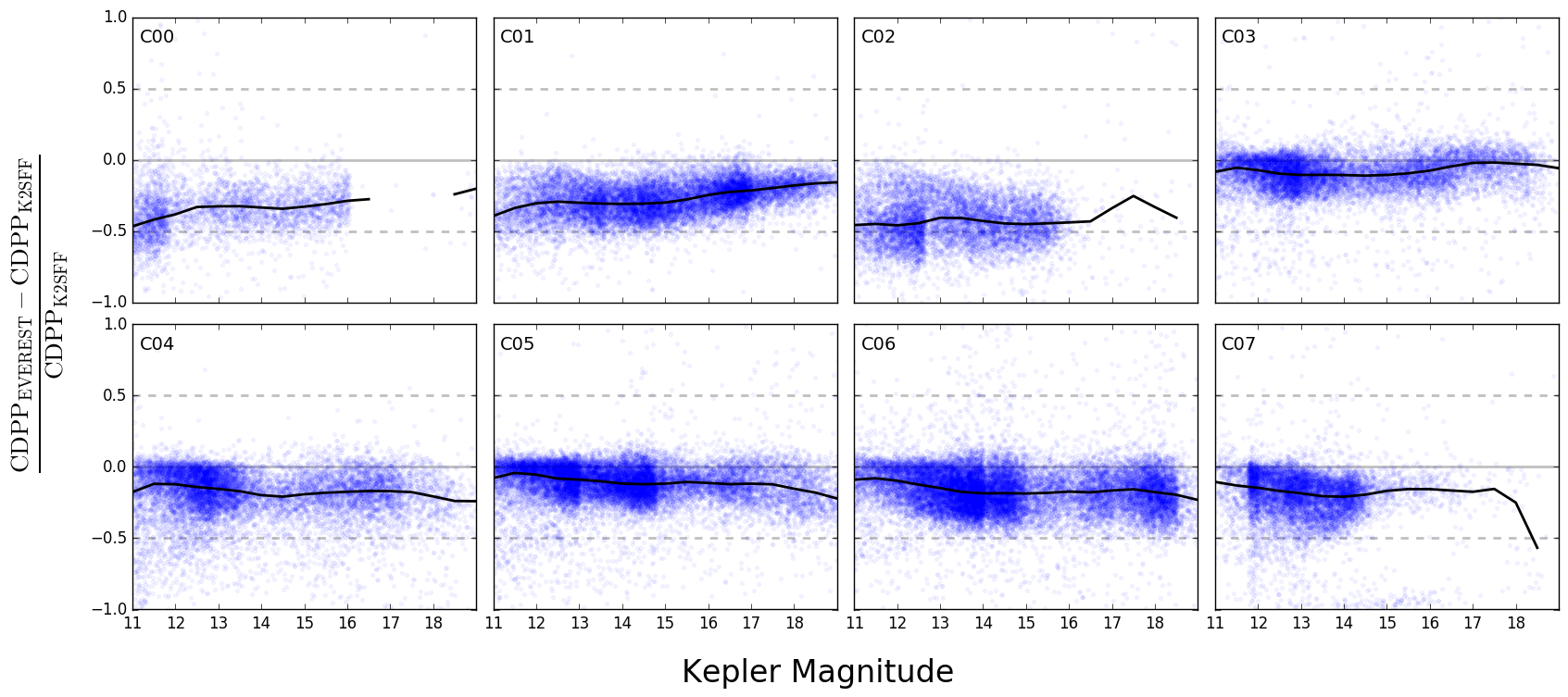,width=7in}
       \caption{A comparison of the \texttt{EVEREST} and \texttt{K2SFF} CDPP for
                each individual campaign. Note the marked difference between
                campaigns 3-7 and campaigns 0-2. For campaign 2, in particular,
                the relative improvement is close to 0.5, corresponding to an average
                \texttt{EVEREST} precision a factor of 2 higher than \texttt{K2SFF}.}
     \label{fig:comparison_k2sff_by_campaign}
  \end{center}
\end{figure*}

\begin{figure}[h]
  \begin{center}
      \psfig{file=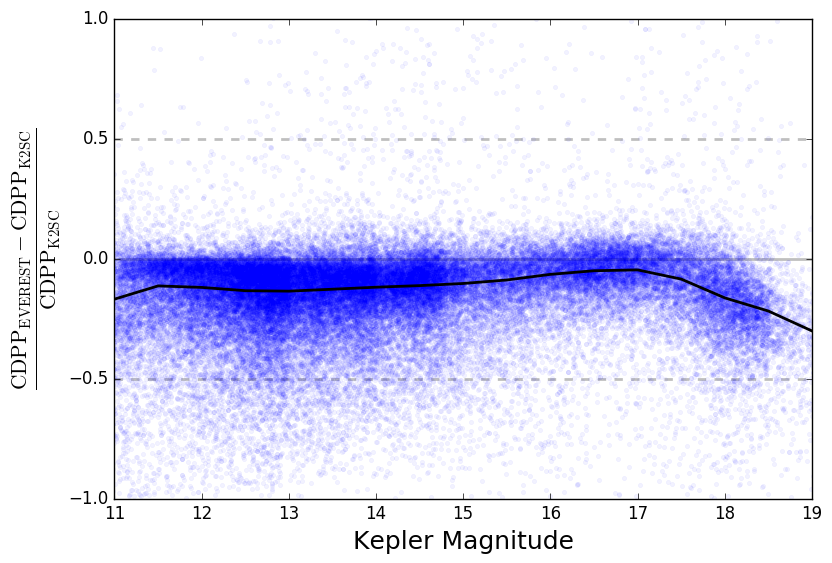,width=3.5in}
       \caption{Same as Figure~\ref{fig:comparison_k2varcat}, but comparing 
                \texttt{EVEREST} to \texttt{K2SC}. To ensure both sets of light curves
                are on the same footing, the \texttt{K2SC} CDPP
                is computed from the PDC flux corrected for the instrumental 
                systematics \emph{only}. As before, a Savitsky-Golay filter
                is then applied to both sets of light curves. The median relative difference
                is once again negative everywhere, indicating that \texttt{EVEREST} yields higher
                precision light curves at all magnitudes.}
     \label{fig:comparison_k2sc}
  \end{center}
\end{figure}

In this section we compare the precision of our de-trended light curves to that of the
\texttt{K2VARCAT} \citep{ARM15}, \texttt{K2SFF} \citep{VJ14}, and \texttt{K2SC} \citep{AIG16} light curves. Though other
pipelines exist \citep[e.g.,][]{LUN15, HUA15, CRO15}, here we focus on those that 
are also available for download at
the MAST HLSP \emph{K2} archive.\footnote{https://archive.stsci.edu/k2/hlsp/}
For each of the pipelines, we download all available de-trended light curves and
compute the CDPP as described in \S\ref{sec:precision}. In 
Figures~\ref{fig:comparison_k2varcat}-\ref{fig:comparison_k2sc}, we plot a pairwise
comparison of the precision achieved for all campaign 0-7 light curves that are
available in both our catalog and that of the other pipeline. These plots are based
on Figure 10 in \cite{AIG16}; the plots show the relative 6-hr CDPP difference between
our light curves and those produced by the \texttt{K2VARCAT}, \texttt{K2SFF}, and \texttt{K2SC}
pipelines as a function of Kepler magnitude. Results for individual light curves are indicated by
blue dots, and the median in half magnitude-wide bins is shown as a black line.
As before, we trim
5$\sigma$ outliers and apply a high-pass filter prior to computing the CDPP.
We again exclude from the comparison saturated stars and stars in highly crowded apertures.

Figure~\ref{fig:comparison_k2varcat} shows the comparison to the \texttt{K2VARCAT}
pipeline. Our de-trended precision is systematically better at all magnitudes shown;
on average, we achieve about double the precision (half the CDPP). There are very
few cases where \texttt{K2VARCAT} yields lower CDPP values than \texttt{EVEREST}
(points above zero).

Figure~\ref{fig:comparison_k2sff} shows the same plot, but comparing \texttt{EVEREST} to
\texttt{K2SFF}. Again, \texttt{EVEREST} yields higher median precision at all
magnitudes; \texttt{EVEREST} light curves have $\sim 20\%$ less scatter on average. The
gain in precision changes with $K_p$, but also with campaign; this can be seen in 
Figure~\ref{fig:comparison_k2sff_by_campaign}. Note, in particular, that the \texttt{EVEREST}
precision is significantly higher relative to that of \texttt{K2SFF} in campaigns 0-2;
in campaign 2, particularly, the average star has half the scatter in the \texttt{EVEREST}
light curves. In the more recent campaigns, the precision gain relative to \texttt{K2SFF}
is smaller, but \texttt{EVEREST} still yields higher median precision at all magnitudes.
In general, \texttt{EVEREST} light curves also have significantly fewer outliers than \texttt{K2SFF} light curves.
This is clear from Figures~\ref{fig:detrended0} and \ref{fig:detrended1} in the
following section: \texttt{K2SFF} light curves often display a band of outliers
above or below the light curve continuum, most likely associated with thruster firings. PLD
naturally corrects for these, obviating the need to discard the several hundred thruster
firing data points in each campaign.

Finally, in Figure~\ref{fig:comparison_k2sc} we plot the comparison to the PDC version
of the \texttt{K2SC}
pipeline, whose performance is the best out of the three we consider here. In order to 
ensure the two datasets are compared on equal footing, we use the systematics-corrected
\texttt{K2SC} fluxes rather than the fully whitened fluxes; we obtain these by
summing the \texttt{FLUX} and \texttt{TREND\_T} columns in the dataset. We then apply
a Savitsky-Golay filter, as we do to the \texttt{EVEREST} data, to remove the stellar 
components of the variability. Since campaigns 0-2 are not available in the \texttt{K2SC}
catalog, this plot shows the comparison for campaigns 3-5 only.
Once again, \texttt{EVEREST} light curves have higher precision than those of
\text{K2SC} by $\sim$ 10\% for bright stars and $\sim$ 5\% for stars in the range
$16 \lesssim K_p \lesssim 11$.

All plots shown here display a significant amount of scatter. While \texttt{EVEREST}
yields the lowest median CDPP in all cases, we recommend comparing light curves from
the different pipelines when the highest precision is desired for specific targets.
Moreover, at this time \texttt{EVEREST} performs poorly for saturated
targets and for those in very crowded fields.


\pagebreak

\subsubsection{Example Light Curves}
\begin{figure*}[t]
  \begin{center}
    \leavevmode
      \psfig{file=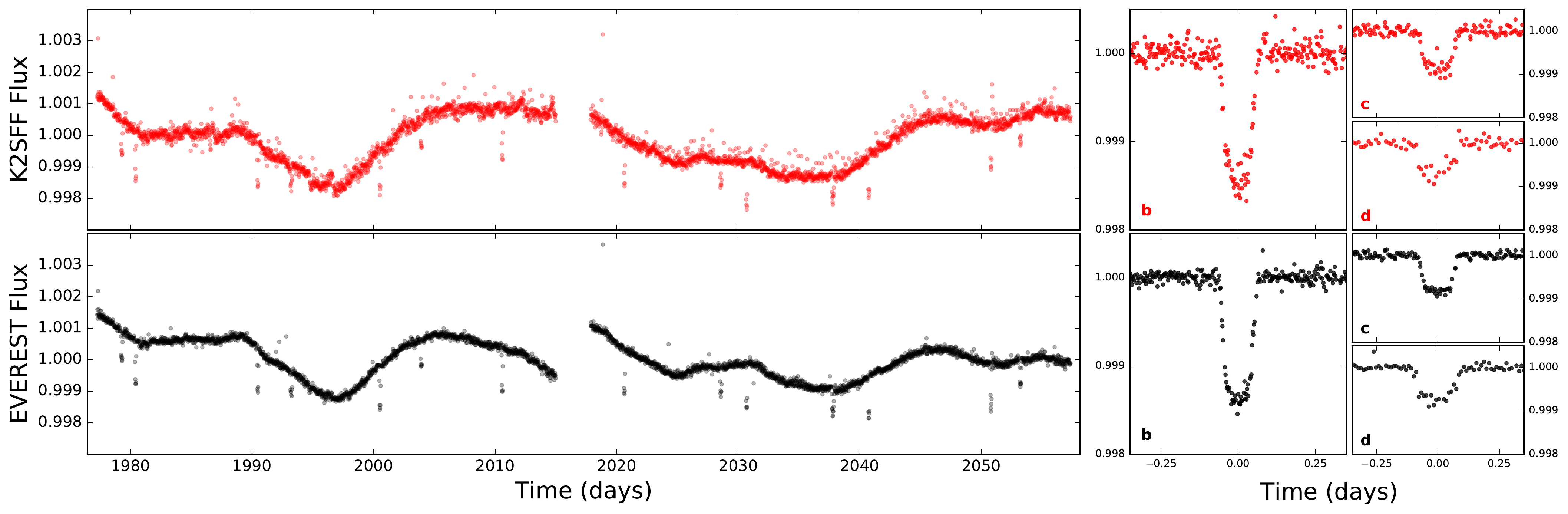,width=7in}
       \caption{De-trended light curves for the campaign 1 star EPIC 201367065 \citep[K2-3,][]{CRO15}. \emph{Top}: The de-trended
                \texttt{K2SFF} flux (left) and the GP-smoothed flux folded on the periods of 
                the planets b, c, and d (right). \emph{Bottom}: The de-trended \texttt{EVEREST}
                flux. The 6-hr CDPP is 30.9 ppm for \texttt{K2SFF} and 16.6 ppm for
                \texttt{EVEREST}, a factor of $\sim$ 2 improvement.}
     \label{fig:detrended0}
  \end{center}
\end{figure*}
\begin{figure*}[t]
  \begin{center}
    \leavevmode
      \psfig{file=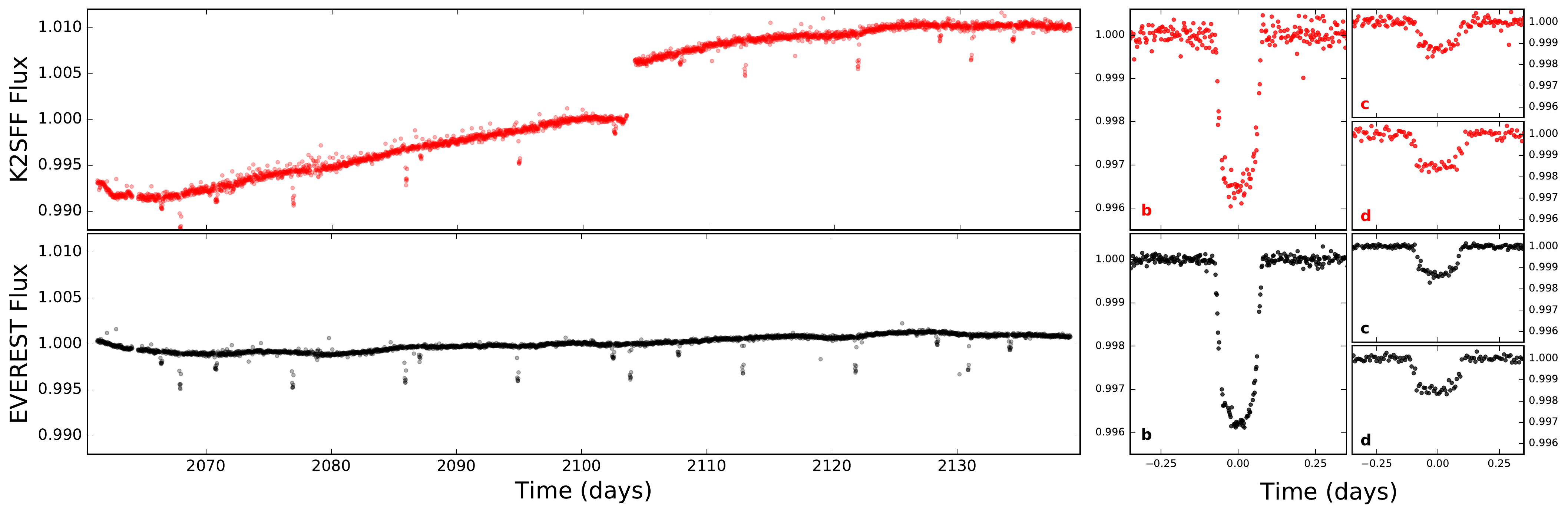,width=7in}
       \caption{De-trended light curves for EPIC 205071984, a campaign 2 star with three
                known planet candidates \citep{SIN15}. As in Figure~\ref{fig:detrended0}, the
                \texttt{K2SFF} light curve and the folded transits of EPIC 205071984.01 (b),
                205071984.02 (c), and 205071984.03 (d)
                are shown at the top; the equivalent plots for \texttt{EVEREST} are shown
                at the bottom. The 6-hr CDPP is 56.1 ppm for \texttt{K2SFF} and 24.0 ppm for
                \texttt{EVEREST}, a factor of $\gtrsim$ 2 improvement.}
     \label{fig:detrended1}
  \end{center}
\end{figure*}
In Figures~\ref{fig:detrended0} and \ref{fig:detrended1} we plot our de-trended light curves
for two \emph{K2} planet candidate hosts, EPIC 201367065 and 205071984. These were chosen
specifically to illustrate the advantages of the PLD technique and are, in a sense, best
case scenarios. Both stars have $K_p \approx 12$ and are the only bright sources in their
apertures. (There are two faint sources, $K_p \approx 17$ and $K_p \approx 18$, 
at the edge of the aperture of EPIC 205071984, but they are faint enough as to not affect
the de-trending.) In each of the figures, we plot the \texttt{K2SFF} light curves at the top (red)
and the \texttt{EVEREST} light curves at the bottom (black). To the right, we plot the folded
transits of their planet candidates after removing the stellar variability signal with a GP.

In both cases, the \texttt{EVEREST} precision is a factor of about 2 higher than that of
\texttt{K2SFF}: \texttt{EVEREST} yields 30.9 ppm for EPIC 201367065 and 24.0 ppm for EPIC 205071984. 
This is visible 
in both the full light curve and in the folded transits,
which display significantly less scatter. Note, importantly, that the greater de-trending
power of \texttt{EVEREST} does not affect the low-frequency stellar variability, which is 
present equally in both sets of light curves.

The corresponding light curves in the \texttt{K2VARCAT} database have CDPP values of 43.1
and 63.4, respectively. At the time of writing, these light curves are not present in the
\texttt{K2SC} catalog.

\section{Conclusions}
\label{sec:conclusions}
In this paper we introduced \texttt{EVEREST}, a pipeline developed to yield the highest
precision light curves for \emph{K2} stars. \texttt{EVEREST} builds on the pixel level
decorrelation (PLD) technique of \cite{DEM15}, extending it to third order in the pixel
fluxes and combining it with principal component analysis to yield a set of basis
vectors that together capture most of the instrumental variability in the data. Gaussian
process (GP) regression is then performed to remove the instrumental
systematics while preserving astrophysical signals. 
In order to prevent overfitting, we
developed a method to determine the ideal number of principal components to use in the fit, yielding reliable,
high precision de-trended light curves for all \emph{K2} campaigns to date.

We validated our model by performing transit injection and recovery tests and showed that
when transits were properly masked by our iterative sigma-clipping technique, we recovered the correct depths
without any bias.
When transits were not masked (the case of many low signal-to-noise transits, which
are missed by our outlier detection step), PLD de-trending
resulted in somewhat shallower transits by $\lesssim 10\%$. 
We therefore strongly encourage those making use of our light curves for transiting exoplanet
characterization to run \texttt{EVEREST} while explicitly masking these transits. Our code
is implemented in Python, is user-friendly, and takes only a few seconds to run for a given
target. The same applies to those using our light
curves for transiting planet searches. Once features of interest are detected, one
should run \texttt{EVEREST} again with those features masked to obtain unbiased
estimates of the transit parameters.
While the decreased transit depth can in principle preclude the detection of very low signal-to-noise
planets in the \texttt{EVEREST} light curves, we find that the increased precision of these light curves
relative to other pipelines is sufficient to enable the 
detection of previously unknown, small transiting planets around \emph{K2} host stars \citep{KRU16}.
Since \texttt{EVEREST} preserves stellar signals,
these light curves should also greatly aid in stellar variability and asteroseismology studies.

For stars brighter than $K_p \approx 13$, we found that \texttt{EVEREST} recovers the 
photometric precision of the original \emph{Kepler} mission; for fainter stars, the 
median precision is within a factor of 2 of that of the original mission.
We further compared our de-trended light curves to those produced by the other \emph{K2} pipelines
available at the MAST HLSP \emph{K2} archive.
\texttt{EVEREST} light curves have systematically higher precision than \texttt{K2SFF},
\texttt{K2VARCAT} and \texttt{K2SC} for all Kepler magnitudes $K_p > 11$. Currently, \texttt{EVEREST}
performs poorly for saturated targets and for those in highly crowded fields.

Our catalog of de-trended light curves is publicly available at 
\texttt{\url{https://archive.stsci.edu/prepds/everest/}} and will be constantly
updated for new \emph{K2} campaigns. The code used to de-trend the light curves
is open source under the MIT license and is available at \texttt{\url{https://github.com/rodluger/everest}},
along with user-friendly routines for downloading and interacting with the de-trended light curves.
A static release of version \texttt{1.0} of the code is also available at 
\texttt{\url{http://dx.doi.org/10.5281/zenodo.56577}}.
Since the only
inputs to \texttt{EVEREST} are the pixel level light curves, the techniques developed
here can be generally applied to light curves produced by any photometry mission,
including the upcoming \emph{TESS} mission \citep{RIC15}, to remove instrumental noise
and enable the detection of small transiting planets.

\begin{acknowledgments}
\small{
R.L., R.B. and E.A. acknowledge support from NASA grant NNX14AK26G
and from the NASA Astrobiology
Institute's Virtual Planetary Laboratory Lead Team, funded
through the NASA Astrobiology Institute under solicitation
NNH12ZDA002C and Cooperative Agreement Number
NNA13AA93A. E.A. acknowledges additional support from NASA grants 
NNX13AF20G and NNX13AF62G. E.K. acknowledges support from an NSF 
Graduate Student Research Fellowship.
This research used the advanced computational,
storage, and networking infrastructure provided by the
Hyak supercomputer system at the University of Washington.}
\end{acknowledgments}

\vfill

\pagebreak

\bibliographystyle{apj}
\bibliography{k2}
\end{document}